\documentclass[]{aa}

\newcommand{\Msun}{M$_\odot$}

\newcommand{\Teff}{$T_{\rm eff}$}

\newcommand{\kms}{km~s$^{-1}$}
\newcommand{\Ha}{H$_\alpha$}

\usepackage{color,graphicx,rotating,natbib}
\begin{document}
\titlerunning{Metal-deficient Ba stars and yellow symbiotics}
\title{On metal-deficient barium stars and their link with yellow symbiotic 
stars\thanks{Based on observations carried out at the European Southern 
Observatory (ESO, La Silla, Chile) and with the 1-m Swiss telescope at
the Haute-Provence Observatory}}

\author{
A. Jorissen\inst{1}\thanks{Senior Research Associate,
F.N.R.S., Belgium}
\and
L. Za{\v c}s\inst{2}
\and
S. Udry\inst{3}
\and
H. Lindgren\inst{4}
\and
F.A. Musaev\inst{5,6,7} 
}
\institute{ 
Institut d'Astronomie et d'Astrophysique, Universit\'e Libre de
Bruxelles, CP 226, Boulevard du Triomphe, B-1050 Bruxelles,
Belgium
\and
Institute of Atomic Physics and Spectroscopy, University of Latvia, 
Rai\c{n}a bulv\= aris 19, R\=\i ga, LV 1586 Latvia
\and
Observatoire de Gen\`eve, CH-1290 Sauverny, Suisse
\and
Lund Observatory, Box 43, SE-221 00, Lund, Sweden
\and
Special Astrophysical Observatory and Isaac Newton Institute of
Chile, SAO Branch, Nizhnij Arkhyz, 369167, Russia 
\and
ICAMER, National Academy of Sciences of Ukraine,
361605 Peak Terskol, Kabardino-Balkaria, Russia
\and
Shamakhy Astrophysical Observatory, National Academy of Sciences
of Azerbaijan, Azerbaijan 
}

\date{Received / Accepted }

\abstract{
This paper addresses the question of why metal-deficient barium stars
are not yellow symbiotic stars (YSyS).  
Samples of (suspected) metal-deficient barium (mdBa) stars and
YSyS have been collected from the literature, and their properties
reviewed. It appears in particular that the barium nature of the
suspected mdBa stars needs to be ascertained by detailed abundance analyses.  
Abundances are therefore derived for two of them, HD~139409 and
HD~148897, which reveal that HD~148897 should not be considered a
barium star.  HD~139409 is a mild barium star, with
overabundances observed only for elements belonging to the first
s-process peak (Y and Zr). It is only moderately metal-poor ([Fe/H] $=-0.4$).
The evidence for binarity among
metal-deficient barium stars is then reviewed, using three different methods: (i) radial-velocity variations (from CORAVEL observations), (ii) Hipparcos astrometric data,  and (iii)  a method based on the 
comparison between the Hipparcos and Tycho-2 proper motions. An orbit
is obtained for HIP~55852, whereas evidence for the (so far unknown)
binary nature of HIP~34795, HIP~76605, HIP~97874 and HIP~107478 is
presented. No conclusion regarding the binary nature of HIP~11595,
HIP~25161 could be reached. Two stars
with no evidence for binarity whatsoever (HIP~58596 and
BD~+3$^\circ$2688) are candidates low-metallicity 
thermally-pulsing asymptotic giant
branch stars, as inferred from their large luminosities.  The reason
why mdBa stars are not YSyS is suggested to lie in their different
orbital period distributions: mdBa stars have on average longer
orbital periods than YSyS, and hence their companion accretes matter
at a lower rate, for a given mass loss rate of the giant star. The
definite validation of this explanation should nevertheless await the
determination of the orbital periods for the many mdBa stars still
lacking periods, in order to make the comparison more significant.
\keywords{binaries: symbiotic - Stars: abundances - Stars: AGB and post-AGB 
-   binaries: spectroscopic } 
             }
\maketitle

\section{The problem}

Our understanding of the link between chemically-peculiar red
giants like barium stars or CH stars, and yellow symbiotic stars (YSyS) 
has made substantial progress in the last decade \citep[see the reviews
  by][]{Jorissen-03a,Jorissen-03}, mainly with the 
realisation that likely all yellow symbiotics (i.e., involving a
giant of spectral type G or K as primary component) involve 
barium stars \citep{Smith-96,Smith-97,Pereira-97,Pereira-98}. 
The metal-deficient
nature of the giant is a key factor, because it implies a rather large
luminosity for the giant. Since evolutionary tracks of
metal-deficient stars are shifted towards the blue,
metal-deficient giants of spectral type K must lie on the upper part
of the (asymptotic) giant branch \citep[see Fig.~11 of ][]{Smith-96}, 
where they suffer strong mass
loss. If such metal-deficient K giants are in binary systems, their
strong wind will interact with the companion and trigger symbiotic
activity.  

The other facet of this problem, namely whether all metal-deficient
barium stars are symbiotic stars, is not yet fully answered. The
present paper offers a first step in that direction.  

We have collected a list of candidate metal-deficient barium stars and
have assembled new observations to check (i) whether these stars are
binaries, (ii) whether they are barium stars and (iii) whether they
exhibit symbiotic activity.

\section{The samples}

Before discussing metal-deficient barium stars, it is useful to
first summarize the properties of YSyS, to which 
metal-deficient barium stars may be compared.

\renewcommand{\baselinestretch}{1}
\begin{table*}
\caption[]{\label{Tab:yellow}
The barium syndrome among YSyS. The spectral type of the
cool component is taken from M\"urset \& Schmid (1999), or references therein.
In column labelled `nebula', `y' means that an optical nebula has been 
detected, and `PN' that, based on its emission line
spectrum, the star has traditionally been
included in planetary nebulae catalogues,
even though no optical nebula may be visible. The column labelled '$P$' lists 
the orbital period, from \citet{Muerset-99}}

\begin{tabular}{lllrrclllll}
\medskip\\
\hline
\smallskip\\
 Name          & Sp. Typ. & [Fe/H]   & \multicolumn{1}{c}{$V_r$}    &
\multicolumn{1}{c}{$b$}   & [Ba/Fe] &
$V \sin \; i$ & nebula & $P$  & Ref. \\

               &          &          & \multicolumn{1}{c}{(km/s)}   
                                         & \multicolumn{1}{c}{($^\circ$)}
&& \multicolumn{1}{c}{(km/s)} & & (d) &\\
\hline
\medskip\\
\noalign{\hspace*{\fill}d'-type\hspace*{\fill}}\\

\hline\\
V417 Cen & G8-K2 & $\sim 0.0$ & & $-1$ & 0.5 &  70 & y &  247 & (5,11)\\
HDE 330036& G5 & 0.02 & $-14$ & $+4$ & 0.88 & 100 & PN & - & (14,17)\\
\multicolumn{1}{r}{=Cn 1-1}\cr 
AS 201  & G5 & 0.07 & & $+7$ & 0.63 &  25 & y & - & (12,17)\\
V471/V741 Per & G5 &  ? &$-$12 & $-$9 & $> 0$ & &PN & - &(2)\\ 
\multicolumn{1}{r}{=M 1-2}\cr 
St H$_\alpha$ 190 & G5 & 0.0 & $\sim 10$ & $-$35 & $\sim 0.5$ & 100 &
bip. outf.& - & (10,13)\\ 
Wray 157 & G5 & ? \\ 
Hen 1591 & $<$ K4 & ?\\
\hline\\
\noalign{\hspace*{\fill}s-type\hspace*{\fill}}\\
\hline\\
UKS Ce-1 & C4,5Jch & ? & +20 & +20 & $>0$ &&&  - & (6)\\
S 32  & C1,1CH & ? & +325 & $-$30 &  $>0$ &  & & 612 & (6,14)\\
Hen 2-467 & K0  & -1.1 & $-$109 & $-$12 &  +0.8  &&n & 478 & (4,16)\\ 
BD$-21^\circ$3873 & K2 & -1.1 & +204 & +37 & +0.5 & & n &282 & (3,15,16) \\
           &     & -1.3   &          &        & +0.3 & &&&(9)\\  
AG Dra & K2 & -1.3 & $-$148 & +41 & +0.5 & & n &554 & (8,16)\\
CD $-43^\circ$14304 & K7 &-1.4 & +27 & $-41$ & ?  &  &&1448 & (7,18)\\
\hline
\medskip\\
\noalign{\hspace*{\fill}Chemical evolution of the Galaxy\hspace*{\fill}}\\
   & & -1.0  & & &   $< 0.2$ &&&  & (1) \medskip\\
\hline
\end{tabular}
\vspace{3mm}\\
{\small References: 
(1) Edvardsson et al., 1993, A\&A, 275, 101 (2) Grauer \& Bond, 1981, PASP,
93, 630 (3) Pereira et al., 1997, AJ, 114, 2128 (4) Pereira et al., 1998, AJ, 116,
1977 (5) Pereira et al., 2003, In: Symbiotic stars probing stellar evolution,
R.L.M. Corradi, J. Miko\l ajewska, T.J. Mahoney (eds.), 
Astron. Soc. Pacific Conf. Ser. (San Francisco), p. 85  
(6) Schmid, 1994, A\&A, 284, 156 (7) Schmid et al., 1998, A\&A, 329, 986 
(8) Smith et al., 1996, A\&A, 315,
179 (9) Smith et al., 1997, A\&A, 324, 97 (10) Smith et al., 2001, ApJ, 556,
L55 (11) Van Winckel et al., 1994, A\&A, 285, 241  (12) Schwarz, 1991, A\&A,
243, 469 (13) Munari et al., 2001, A\&A, 369, L1 
(14) Schmid \& Nussbaumer, 1993, A\&A, 268, 159
(15) Munari \& Patat, 1993, A\&A, 277, 195 
(16) Corradi et al., 1999, A\&A, 343, 841
(17) Pereira et al., 2005, A\&A 429, 993
(18) The metallicity is from Pereira, priv. comm.
}
\end{table*}

\subsection{Yellow symbiotic stars}

\begin{figure*}
\resizebox{0.49 \hsize}{!}{\includegraphics{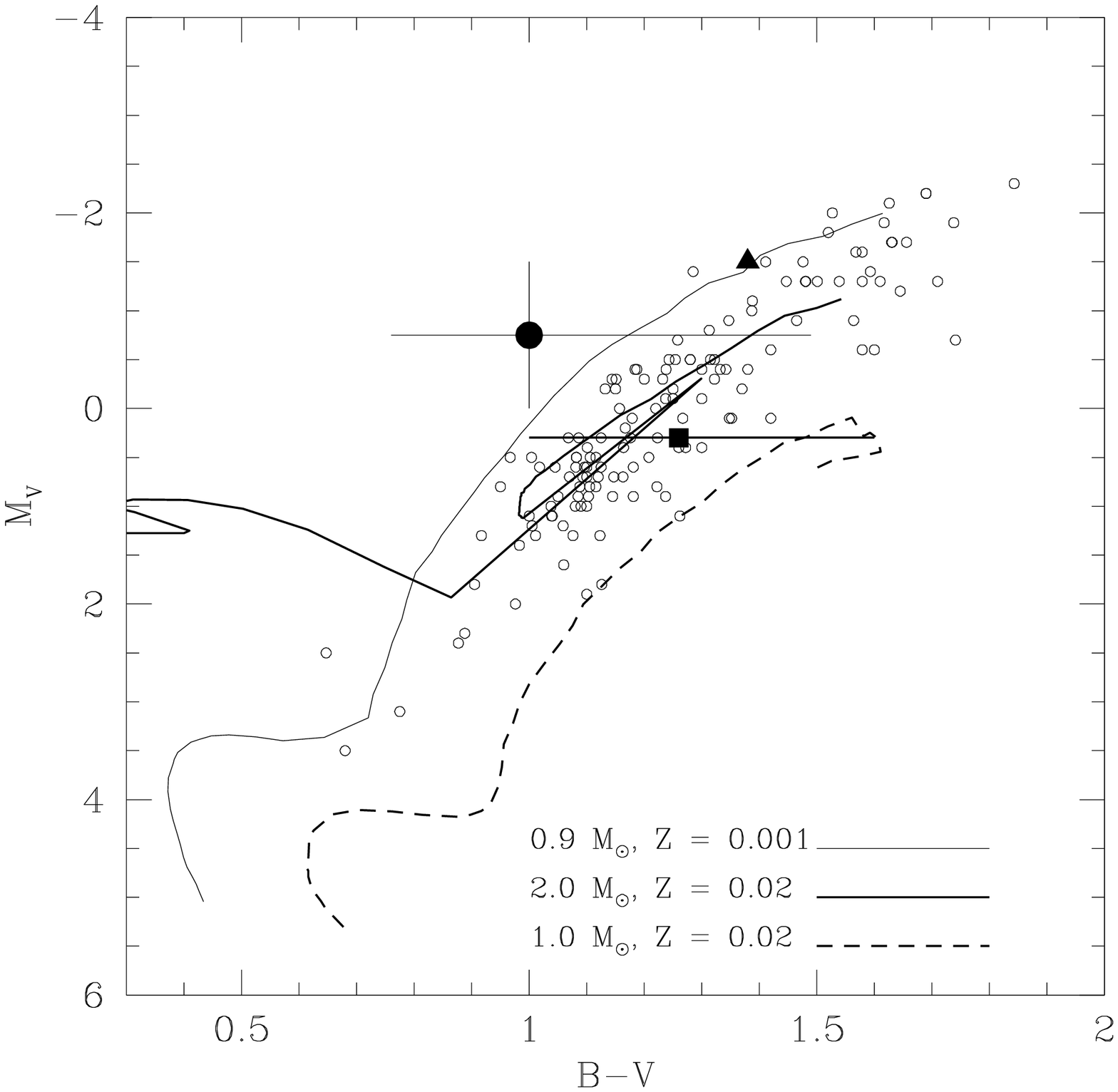}}
\resizebox{0.49 \hsize}{!}{\includegraphics{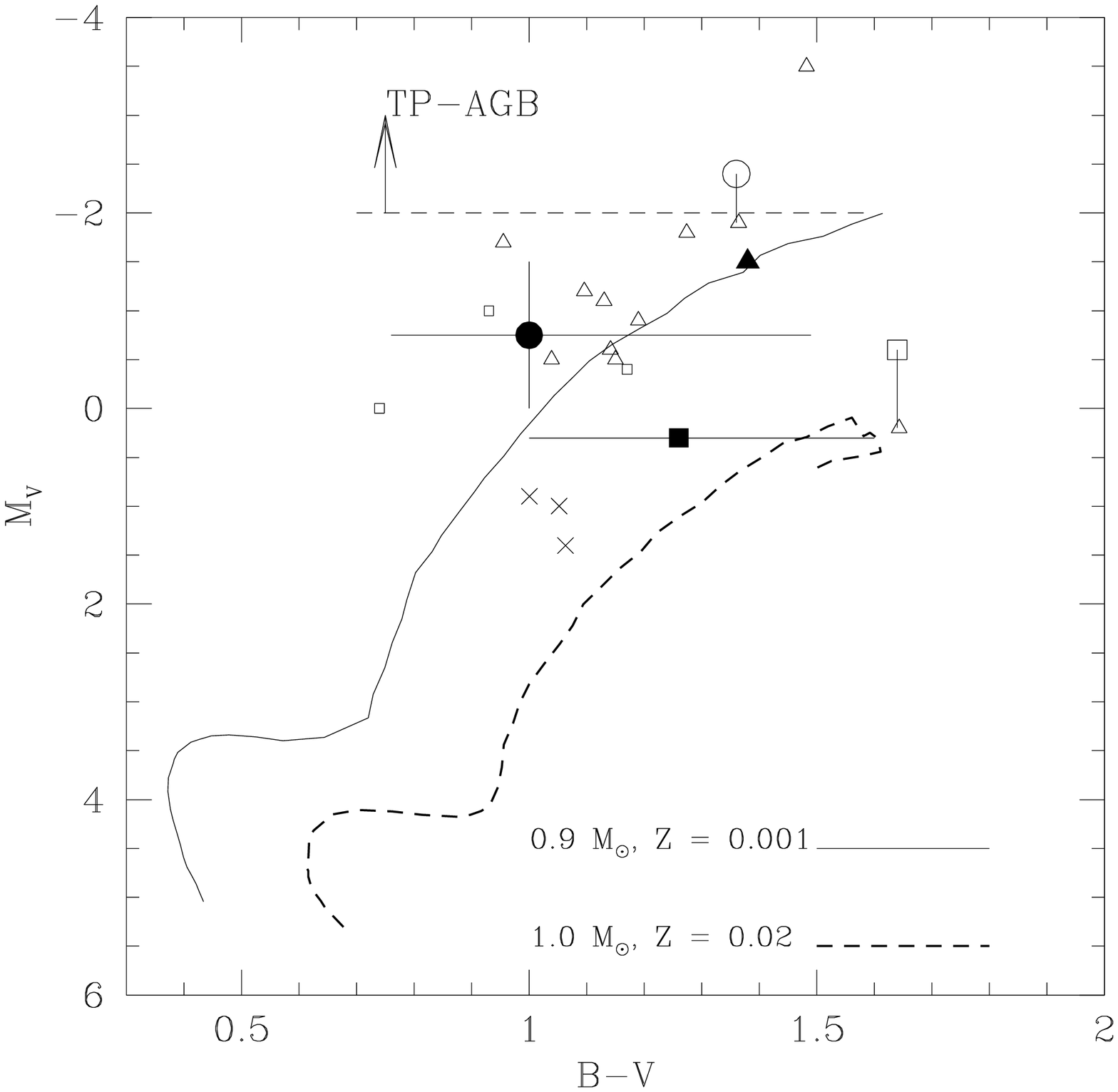}}
\caption[]{\label{Fig:mdBa}
Left panel (a): Evolutionary tracks of \protect\cite{Schaller-92} compared
with the locations of classical barium stars (stars labelled `G' in 
\protect\citet{Mennessier-97}; open dots) and the yellow SyS AG~Dra
(filled circle), BD $-21^\circ3873$ (filled triangle) and Hen 2-467 (filled 
square). The bolometric magnitudes were taken from the references
listed in Table~\ref{Tab:yellow}. These bolometric magnitudes were combined
with bolometric corrections from \protect\citet{Bessell-98} and 
$B-V$ indices from \protect\citet{Munari-92} and \protect\citet{Munari-Buson-92} to yield the absolute visual magnitudes.\\ 
Right panel (b): Same as (a) but
for YSyS (filled symbols as in the left panel) and metal-deficient
barium stars [open triangles: 
stars flagged as `H' by \protect\citet{Mennessier-97}; crosses: CH stars
also flagged as `H' by \protect\citet{Mennessier-97}; small open squares:
additional metal-deficient, s-process-rich stars 
from Table~\protect\ref{Tab:H}; 
open circle: HD 104340, large open square: HD 206983 from
\protect\citet{Junqueira-2001}].  
The dashed horizontal line represents the luminosity ($M_{\rm bol} =
-3$, corresponding to  $M_{\rm V} \sim -2$) 
at the first thermal pulse in a 1~\Msun\ AGB star of metallicity [Fe/H] =$-1.8$
according to \protect\citet{Lattanzio-91}.
}
\end{figure*}

All known YSyS are listed in Table~\ref{Tab:yellow}, which 
shows that all the stars
studied so far exhibit the barium syndrome.
YSyS with a {\it stellar} infrared continuum (s-type, as opposed 
to the dusty d'-type; see below) 
are clearly metal-deficient objects, as revealed by their low
metallicities and high space velocities (CD $-43^\circ14304$ may be an exception; 
however, it is of spectral type K7, and should perhaps not be included
in the family of YSyS). The presence of the barium
syndrome among a family of binary metal-deficient stars fully supports
the commonly accepted hypothesis that the s-process is more efficient
at low metallicities \citep{Clayton-1988,Jorissen-03a}.
s-Type YSyS, 
with their metallicities lower than classical barium stars, may be
expected to be, on average, more luminous than the latter \citep[see Fig.~11
of][comparing the luminosity function of Pop.I and
Pop.II K giants]{Smith-96}.   This is a direct consequence of
the fact that evolutionary tracks
shift towards the blue in the Hertzsprung-Russell (HR) diagram as metallicity
decreases, as shown in Fig.~\ref{Fig:mdBa}b.
Fig.~\ref{Fig:mdBa}a confirms that the
YSyS AG~Dra and BD $-21^\circ3873$ are indeed 
more luminous than classical barium stars.
This difference in the average luminosity -- and hence mass-loss rate 
-- of the two populations thus
explains why YSyS, despite hosting a K giant, exhibit symbiotic
activity whereas barium stars do not. The larger mass-loss rates for
the cool components of s-type YSyS  -- as compared to Ba stars
-- may be inferred from the comparison of their IRAS [12] $-$ [25]
color indices, which reflect the amount of dust present in the
system: ([12] $-$ [25])$_{\rm Ba} < $ 0.1, as compared to 0.45 for
AG~Dra \citep{Smith-96}. \citet{Muerset-91} and \citet{Drake-87} 
provide direct measurements (or upper limits) for the mass
loss rates of AG~Dra and of Ba stars, respectively, which confirm the 
above conclusion.

YSyS with a {\it dusty} infrared continuum \citep[d'-type; ][]{Allen-82, Schmid-Nussbaumer-93} differ
from their s-type counterparts in several respects
(Table~\ref{Tab:yellow}): they host a complex
circumstellar environment (including cool dust, 
bipolar outflows, extended optical nebulae or
emission-line spectra closely resembling those of planetary nebulae), the cool components have early spectral types (F to early
K), they are often
fast rotators \citep[with the possible exception of M 1-2 =V471 Per; ][]{Grauer-Bond-81} and, finally, they 
belong to the galactic disk unlike s-type YSyS which belong to the halo.

All these  arguments  suggest that the hot component in d'-type SyS
has just evolved from the AGB to the WD stage. The rather cool dust 
\citep{Schmid-Nussbaumer-93} is a relic from the mass lost by the AGB
star. The optical nebulae observed in d'-type SyS are most likely 
genuine planetary nebulae rather than the 
nebulae associated with the ionized wind of the cool component
\citep{Corradi-99}. 
This is especially clear for AS 201
which actually hosts {\it two} nebulae \citep{Schwarz-91}: 
a large fossil planetary nebula detected by direct imaging, and a small
nebula formed in the wind of the current cool component. Finally, the
rapid rotation of the cool component has likely been caused by spin
accretion from the
former AGB wind like in WIRRING systems  \citep{Jeffries-Stevens-96,
  Jorissen-03}.  The fact that the cool star has
not yet been slowed down by magnetic braking is another indication
that the mass transfer occurred fairly recently \citep{Theuns-96}. 
\citet{Corradi-1997} obtained 4000~y for the age of the nebula around AS~201, 
and 40000~y for V417~Cen. 

\subsection{Metal-deficient barium stars} 

Metal-deficient barium stars (with
metallicities in the range $-1.1$ to $-1.8$, comparable to that of
YSyS) were identified by \citet{Luck-Bond-91},
\citet{Mennessier-97} and \citet{Zacs-2000}, and occupy the same region of the HR diagram 
as YSyS (Fig.~\ref{Fig:mdBa}b). 
The question thus arises why metal-deficient
barium stars are not SyS. Different answers must be sought, depending 
upon their absolute visual magnitudes $M_{\rm V}$.
The most luminous systems, with $M_{\rm V} < -2$, are likely located
on the thermally-pulsing AGB, 
so that their Ba syndrome may be explained by internal
nucleosynthesis. They thus should not be binaries, and therefore cannot 
be SyS! HD~104340 (open circle in Fig.~\ref{Fig:mdBa}b), 
a metal-deficient Ba star studied by \cite{Junqueira-2001}, and
BD~+$03^\circ 2688$ (Table~\ref{Tab:H}) provide good illustrations of this
situation, since  they both lie above the TP-AGB threshold and 
CORAVEL radial-velocity measurements spanning several years do
not reveal any clear orbital motion (Figs.~\ref{Fig:104340_RV} and
\ref{Fig:bd03_2688}, as well as Sect.~\ref{Sect:Binary}). 

The less luminous and warmest among metal-deficient Ba stars, clumping
around $M_{\rm V} \sim +1$ in the HR diagram, 
are also sometimes 
classified as CH stars (crosses  in Fig.~\ref{Fig:mdBa}b).  
They are not losing mass at a large enough
rate to trigger any symbiotic activity, as revealed by their small
[12] $-$ [25] color indices  \citep[$< 0.3$;][]{Smith-96}.

Finally, at intermediate luminosities ($-2 \le M_{\rm V} \le
+1$) where YSyS are located, metal-deficient Ba stars are not luminous
enough to be TP-AGB  (hence they should be binaries), but yet their mass
loss rates must be large enough to trigger symbiotic activity (provided
that the orbital separation is not too large, since it is the mass {\it
accretion} rate by the compact companion which is in fact the key
parameter; see Sect.~\ref{Sect:discussion} and Jorissen 2003a).  It is
thus of great interest to check (i) the Ba nature of those
metal-deficient stars with intermediate luminosities, (ii) their binary
nature, and (iii) their suspected symbiotic activity.  The first two
issues are addressed in Sects.~\ref{Sect:abundances} and
\ref{Sect:Binary}, respectively.

\begin{figure}
\resizebox{\hsize}{!}{\includegraphics{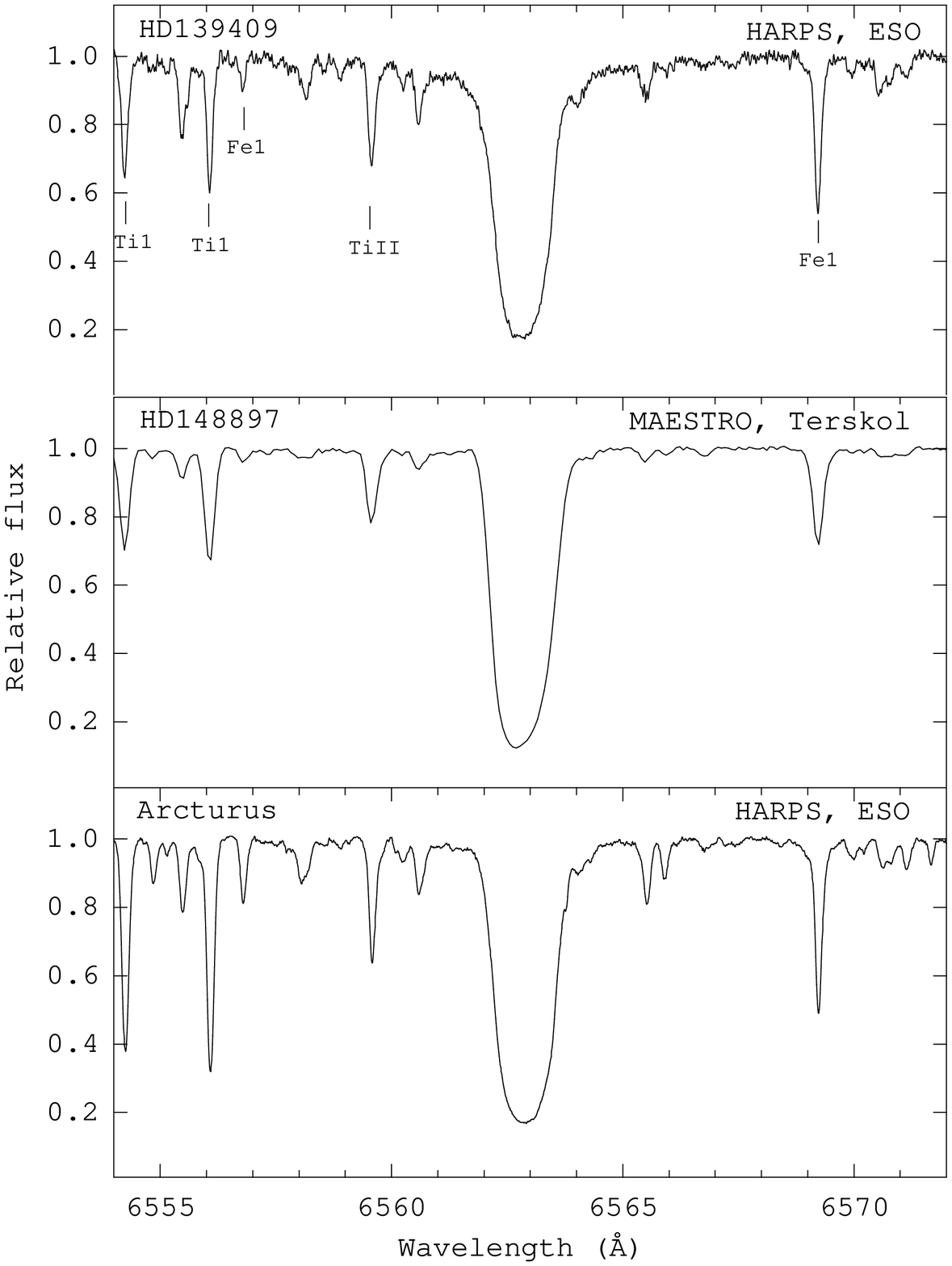}}
\caption[]{\label{Fig:Halpha}
$H_{\alpha}$ profiles for two candidate metal-deficient Ba stars 
(HD~139409 and HD~148897; see however the discussion in 
Sect.~\protect\ref{Sect:abundances}), 
compared to that of 
the K giant Arcturus. 
The wavelength scale has been corrected for
the stellar radial velocity. The profiles show no indication of 
symbiotic activity.  
}
\end{figure}

As far as a possible symbiotic activity is concerned, there is no
indication from their photometric $U-B$ and $B-V$ indices that the
metal-deficient stars in Table~\ref{Tab:H} have a strong blue
continuum which could betray their symbiotic nature. It is thus very
likely that none among these stars is a
full-fledged symbiotic star. 
No signature of weak symbiotic activity \citep[of the kind exhibited by some 
binary S stars; see
  Fig.~\protect\ref{Fig:P} and ][]{VanEck-Jorissen-02} 
is observed in the \Ha\ line profile either
(Fig.~\ref{Fig:Halpha}).  

\begin{figure}
\resizebox{\hsize}{!}{\includegraphics{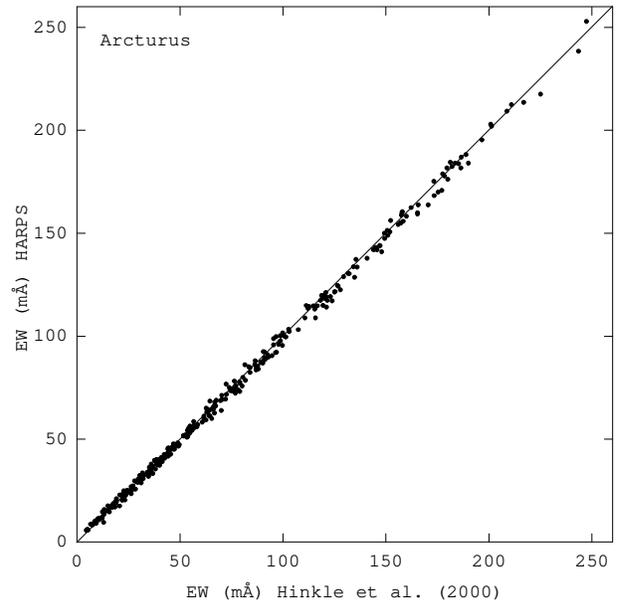}}
\caption[]{\label{Fig:Arcturus}
Comparison of equivalent widths measured on the HARPS spectrum
of Arcturus with those of the Arcturus spectral atlas 
\protect\citep{Hinkle-2000}, for the same set of lines as that measured 
in HD~139409.  
}
\end{figure}

\begin{figure}
\resizebox{\hsize}{!}{\includegraphics{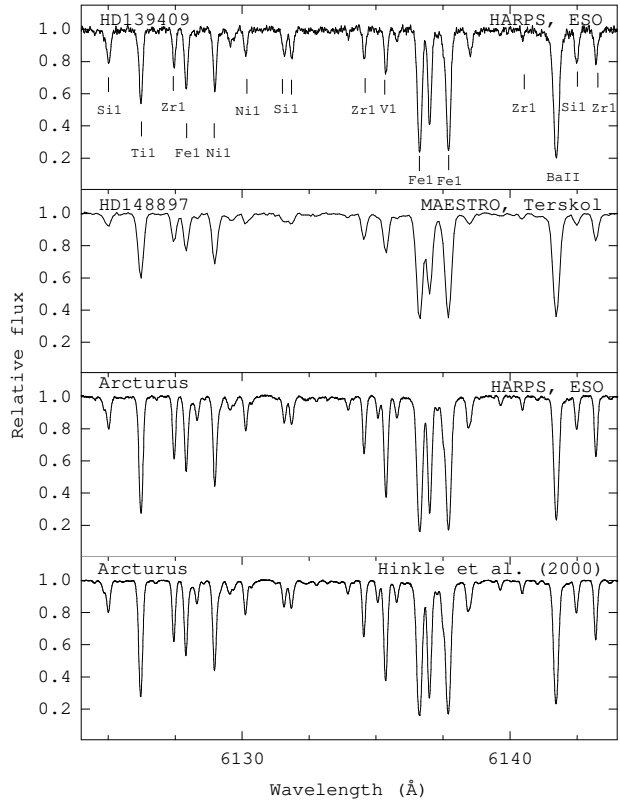}}
\caption[]{\label{Fig:spect_148897}
The spectra of the target stars in the region of the Ba~II line at
614.172~nm. Measured lines are marked. Also shown is the spectrum of
the standard star Arcturus \protect\citep{Hinkle-2000}. 
}
\end{figure}

\begin{figure}
\resizebox{\hsize}{!}{\includegraphics{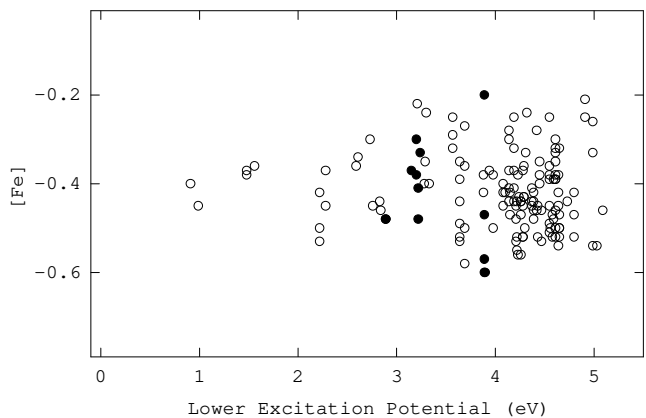}}
\caption[]{\label{Fig:FEvsCHI}
The iron abundances derived from Fe~I (open circles) and Fe~II (filled
circles) lines are displayed 
as a function of the excitation potential for the lower
energy level of the line. The absence of a trend in these data is
used to derive the spectroscopic temperature of 
HD~139409. 
}
\end{figure}

\begin{figure}
\resizebox{\hsize}{!}{\includegraphics{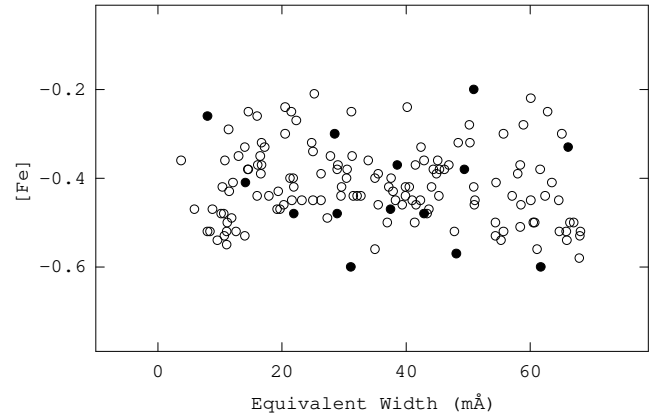}}
\caption[]{\label{Fig:FEvsEW}
The abundances derived from the Fe~I (open circles) and Fe~II (filled
circles) lines are displayed as a function  of the line equivalent widths. 
The absence of a trend in these data is used to derive 
the microturbulent velocity of HD~139409. 
}
\end{figure}

\begin{figure}
\resizebox{\hsize}{!}{\includegraphics{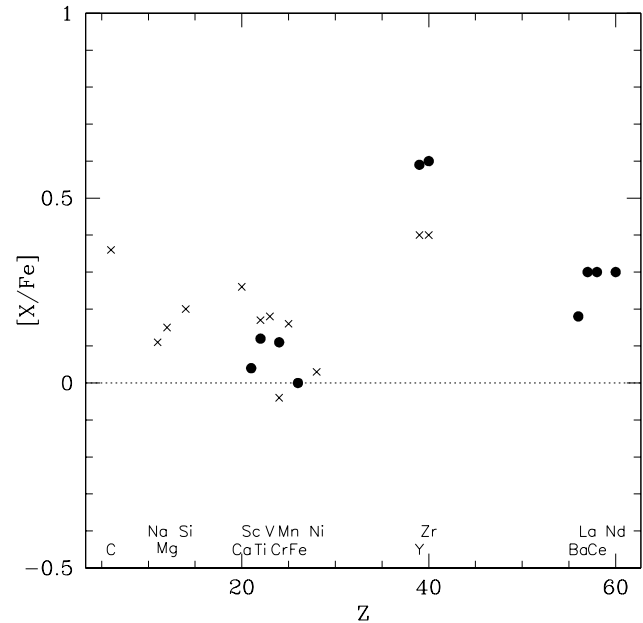}}
\caption[]{\label{Fig:139409}
Abundance pattern in HD~139409 
(abundances derived from ionized species are represented by black dots, and 
from neutral species by crosses).}
\end{figure}

\begin{table*}
\caption[]{\label{Tab:H}
Stars classified as metal-deficient barium stars by
\protect\citet{Mennessier-97}. The absolute magnitude $M_V$ is a
maximum likelihood estimate obtained by \protect\citet{Mennessier-97},
except for the additional stars where it is derived from a straight
inversion of the Hipparcos parallax (BD $+75^\circ348$), from a spectroscopic
estimate of gravity and an educated guess for the mass (BD+$3^\circ 2688$) or from
a fit to the M 92 isochrone (CS stars). The column labelled `Ba' indicates
whether detailed chemical analyses have confirmed the Ba nature of the
star.  }
\begin{tabular}{rrrllllllllll}
\hline
 HIP  &  HD/DM  &  $M_V$  &  $B-V$ & [Fe/H] & Ref & Ba & Ref & Rem\cr
\hline
  4347 & 5424  & -0.6 &1.14 &  &   &   &    \cr
 11595 & 15589 & -0.5 &1.15 & $-0.7$ & 9 \cr
 25161 & $-27^\circ$2233  & -1.1 &1.13 \cr
 29740 &43389  & -3.5 &1.48 &  &   &   &   \cr
 34795 &55496  & -1.7 &0.96 & $-1.55$ & 5 & y & 5 \cr
58596 &104340 & -1.9 &1.36 & $-1.72$ & 3 & y & 3 & \cr
 69834 &123396 & -0.9 &1.19 \cr
 76605 &139409 & -0.5 &1.04 & $-0.42$ & 2 & mild & 2 & $M_V = 1.5$ is
derived from spectroscopic $\log g$  \cr
 80843 &148897 & -1.8 &1.27 & $-1.0$ & 2 & {\bf n}  & 2 & \cr
       &       &      &     & $-0.62$& 7 \cr
       &       &      &     & $-1.16$& 8 \cr
 97874 &187762 & -1.2 &1.10 \cr
107478 &206983 & 0.2 &1.64 & $-1.43$ & 3 & y & 3 & \\
\hline
\medskip\\
\noalign{Additional stars [from refs. (1), (5) and (10)]}
\medskip\\
\hline
43042 &+$75^\circ 348$ & $-0.4:$ & 1.17  & $-0.8$ & 1 & y & 1 \cr 
55852 & +$4^\circ 2466$ & ? & 0.60  & $-1.85$ & 5,11 & y & 4,5 \cr
      & +$3^\circ 2688$ & $-5.0$ & 1.22  & $-1.42$ & 5,11 & y & 5    \cr
      &CS 22942-019 & 0    & 0.74 & $-2.67$ & 10 &  y & 10 \cr
      &CS 22948-027 & $-1$ & 0.93 & $-2.60$ & 10 &  y & 10 \cr
\hline
\end{tabular}

References to the Table:\\
(1) \citet{Zacs-2000} (2) This work (3) \citet{Junqueira-2001}
(4) \citet{Burris-2000} (5) \citet{Luck-Bond-91} (6)
\citet{Jorissen-VE-98} (7) \citet{Kyrolainen-86}  (8)
\citet{Luck-91}
(9) \citet{Barbuy-Jorissen-92}  (10) \citet{Preston-Sneden-01}  (11)
$B-V$ from the Tycho-2 catalog \citep{Hog-00}   
\end{table*}

\section{Abundances}
\label{Sect:abundances}

The classification of the
stars in Table~\ref{Tab:H} as metal-deficient Ba stars is subject to
caution, as it does not rely on spectroscopic abundance analyses, but
rather on a maximum-likelihood assignment based on kinematic, spatial
and luminosity properties \citep{Mennessier-97} for barium stars from the list of \citet{Lu-91}. Nevertheless, when
a metallicity determination is available, it confirms the
metal-deficient nature of the object (see Table~\ref{Tab:H}). HD~139409 is an exception, though, since 
detailed spectral analyses reveal that it is neither metal-poor nor strongly enriched in s-process elements (see below). It may nevertheless
 be hoped that the metal-deficient assignment made by
\citet{Mennessier-97} is valid in all the other cases. Regarding the Ba
nature of these stars, it is known that the \citet{Lu-91} catalogue of barium stars,
from which  the sample of barium stars used by \citet{Mennessier-97} was
drawn, is contaminated by many non-barium stars
\citep{Griffin-Keenan-92,Jorissen-96},  especially among those stars
having a Ba index smaller than 1.  

It would therefore in principle be  necessary to re-evaluate  the Ba nature of all the stars
listed in Table~\ref{Tab:H}. So far, spectra could be obtained for two of
them, HD~139409 and 
HD~148897, which are discussed in detail in the present section.  
Of these, only HD~139409 appears to be a mild barium star, thus confirming
the suspicion about the \citet{Lu-91} catalogue expressed above.

\subsection{The case of HD 148897 and HD 139409}

HD~148897 (= HR 6152) has been tagged as a `likely marginal barium star'
by \citet{Boyle-McClure-75} and as a marginal CH star by
\citet{Vilhu-77}. It therefore found its way into the barium-star catalogs
of \citet{Lu-83} and \citet{Lu-91}, as well as the
catalog of CH and metal-deficient Ba stars of \cite{Sleivite-90}.
The star was classified as G8.5III CN-2 Fe-1 CH-1 by
\cite{Keenan-McNeil-89},
and this classification as CN- and CH-weak contradicts the earlier assignments.

A detailed abundance analysis has thus been performed to clarify the
situation, and its results are compared with previous analyses by 
\citet{Kyrolainen-86} and
\citet{Luck-91} in Sect.~\ref{Sect:results}.

HD~139409 (= HIP~76605) has been classified as a marginal barium star
by \citet{MacConnell-72},  
as G5\,III Ba1 by \citet{Yamashita-81} and as K0\,III/II Ba~0.5 by
\citet{Lu-91}.  

\subsubsection{Observations}

A high-resolution spectrum of HD~148897 was obtained using the Coud\'e 
Matrix Echelle Spectrometer \citep[MAESTRO;][]{Musaev-99} delivering a
resolving power of 45\ts000 and installed on the 2-m Zeiss telescope
of the Terskol Observatory (located in Northern Caucasus at an altitude
of 3100~m). The spectrometer is equipped with a Wright Instruments 
CCD detector with 1242
$\times$ 1152 pixels (22.5 $\times$ 22.5
$\mu$m). A total exposure of 1800~s was taken on February 
18, 2003. The spectrum covers the range 365 to 1020~nm spread over 85
spectral orders. 

A high-resolution spectrum of HD~139409 was obtained on the HARPS spectrograph 
\citep{Mayor-2003}, 
delivering a resolution of 115\ts000 and installed  
on the ESO La Silla~3.6~m telescope. A total
exposure of 200~s was obtained by Xavier
Bonfils on May 31, 2004. In
order to check equivalent widths delivered by HARPS, a spectrum of
the standard star Arcturus was obtained as well by Fabien Carrier. 
The HARPS spectra were reduced by the observers using standard
pipeline processing. Equivalent widths for the same set of lines as those 
studied in HD~139409  
have been measured by one of us (L.Z.) in the HARPS 
spectrum of Arcturus and compared to
those from the Arcturus spectral atlas \citep{Hinkle-2000}. The
agreement between the two sets of equivalent widths is excellent 
(Fig.~\ref{Fig:Arcturus}), thus qualifying 
HARPS for abundance analyses. 
Spectra around the 
$\lambda614.172$~nm Ba~II line are shown in
Fig.~\ref{Fig:spect_148897} for the two analyzed stars and for  
Arcturus \citep{Hinkle-2000}.

\subsubsection{Reduction and analysis}
\label{Sect:results}

{\it HD 148897}

\noindent The reduction of the CCD frames (subtraction of bias, dark and 
scattered light, flat fielding, extraction of echelle orders and
wavelength calibration) was performed with the DECH20T software
\citep{Galazutdinov-92}. More than 500 weak to medium-strong atomic
lines, free of blends, were identified and their equivalent widths were
measured with the DECH routines. The radial velocity was measured using a
large number of symmetric absorption lines. The observed velocity has been
brought to the heliocentric system by adding +22.7~km~s$^{-1}$. The mean
heliocentric radial velocity for HD~144897 was found to be
+16.7~km~s$^{-1}$. The atmospheric parameters for that star 
cannot be derived from
photometry, since the standard temperature calibrations refer to stars
of normal chemical composition. To obtain a colour-independent estimate of
the temperature, a spectroscopic temperature has been derived
from the excitation equilibrium  of Fe~I lines.
The surface gravity $\log g$ was determined using the FeI/FeII ionization
balance, whereas the microturbulent velocity $v_t$ was derived by forcing the
abundances of individual Fe~I lines to be independent of the
reduced equivalent width. The resulting atmospheric parameters for
HD~148897 are as follows: $T_{\rm eff} = 4350$~K, $\log g = 1.0$ (cgs),
and $v_t = 2.0$~\kms. An independent determination of the surface gravity,
using the above \Teff\ value, \citet{Mennessier-97} absolute visual magnitude 
($M_V = -1.8$), a bolometric correction of $-0.5$ and a mass of 1~\Msun, 
yields $\log g = 1.1$, 
in agreement  with the adopted value \citep[see ][for a discussion of the discrepancy
usually observed between the photometric and spectroscopic gravities]{Luck-91}.
Comparison with atmospheric parameters from the literature is presented in
Table~\ref{Tab:abundances}. 

The abundance analysis has been performed with the standard LTE line
analysis program WIDTH9 developed by Kurucz. The model atmospheres
were taken from \citet{Gustafsson-75}. The synthetic spectra
were generated using the spectral synthesis code STARSP
\citep{Tsymbal-96}. Oscillator strengths have been taken from the 
VALD database \citep{Piskunov-95}. 
The resulting abundances 
\citep[normalized by the solar-system abundances of ][]{Grevesse-1998}
are listed in Table~\ref{Tab:abundances}, from which it may be concluded 
that HD~148897 appears to be a rather typical metal-deficient
star \citep{McWilliam-1997}, 
and should certainly {\it not be considered as a (metal-deficient) barium
star}.

Interestingly enough, there are no indications whatsoever from the 
Hipparcos and Tycho data (applying the methods described in
Sect.~\ref{Sect:Binary}) that this star is binary, in agreement with
the fact that it is not a barium star.

\begin{table*}
\caption[]{\label{Tab:abundances}
Averaged absolute abundances (in the scale where $\log \epsilon$(H) $ =
12$) and normalized  abundances [X/Fe] for HD~148897 \citep[relative to the
Sun;][]{Grevesse-1998}. The standard deviations
$\sigma$  and the number of lines used in the analysis ($n$) are also
given. Two sets of abundances from \protect\citet{Luck-91} are given:
one is based on a `photometric $\log g$' (derived from the absolute
magnitude, the effective temperature and an estimate for the stellar
mass) and the other on the `spectroscopic $\log g$' (Fe~I/Fe~II
ionization balance). 
}
\begin{tabular}{lllrllllllllllllllll}
\hline
         &  \multicolumn{5}{c}{This work}  &&
            \multicolumn{3}{c}{Kir\"ol\"ainen et al.  (1986)} &
            \multicolumn{3}{c}{Luck (1991) }  & 
            \multicolumn{3}{c}{Luck (1991) }  \cr
\cline{2-6}\cline{8-9}\cline{11-12}\cline{14-15}
         & \multicolumn{5}{c}{\Teff = 4350~K} && 
           \multicolumn{3}{c}{\Teff = 4360~K} & 
           \multicolumn{3}{c}{\Teff = 4100~K} & 
           \multicolumn{3}{c}{\Teff = 4100~K}  

\cr 
         & \multicolumn{5}{c}{$\log g = 1.0$} &&
           \multicolumn{3}{c}{$\log g = 1.5$} &
           \multicolumn{3}{c}{phot. $\log g = 1.6$} &
           \multicolumn{3}{c}{spec. $\log g = 0.1$} 
\cr
        & \multicolumn{5}{c}{$v_t = 2.0$ \kms}  &&             
          \multicolumn{3}{c}{$v_t = 2.6$ \kms}  &   
          \multicolumn{3}{c}{$v_t = 1.7$ \kms}  &   
          \multicolumn{3}{c}{$v_t = 1.7$ \kms}  
\cr
\cline{2-6}\cline{8-9}\cline{11-12}\cline{14-15}
X	& $\log \epsilon$(X) & $\sigma$ &  $n$ & [X/H]$_\odot$ &
$\left[\frac{{\rm X}_{I,II}}{{\rm Fe}_{I,II}}\right]_\odot$       && [X/H]$_\odot$ & [X/Fe]$_\odot$
&& [X/H]$_\odot$ & $\left[\frac{{\rm X}_{I,II}}{{\rm Fe}_{I,II}}\right]_\odot$  && [X/H]$_\odot$ &
 $\left[\frac{{\rm X}_{I,II}}{{\rm Fe}_{I,II}}\right]_\odot$\cr
\hline
O I  &8.22&0.20    & 4 & 	-0.61 &+0.41&&-0.58 &+0.04 &&-0.50 &+0.50 &&-1.10 &+0.06 \cr 
Na I &5.10&0.08    & 6 & 	-1.23 &-0.21&&-1.11 &-0.49 &&-1.36 &-0.36 &&-1.26 &-0.10 \cr  
Mg I &6.92&0.09    & 6 & 	-0.66	&+0.36&&-0.59 &+0.03 &&-0.54 &+0.46 &&-0.62 &+0.54 \cr
Al I &5.61&0.23    & 4 & 	-0.86	&+0.16&&-0.70 &-0.08 &&-1.28 &-0.28 &&-1.20 &-0.04 \cr 
Si I &6.91&0.11   & 23 & 	-0.77	&+0.25&&-0.56 &+0.06 &&-0.24 &+0.76 &&-0.59 &+0.57 \cr
K I  &4.17&       & 1  & 	-0.95	&+0.07&&-     &-     &&-     &   -  && -    & -    \cr
Ca I &5.65&0.11   & 21 & 	-0.71	&+0.31&&-0.54 &+0.08 &&-1.09 &-0.09 &&-0.97 &+0.19 \cr
Ca II &5.48    &   & 1 & 	-0.88	&+0.09&&-     &-     &&-     &   -  && -    & -    \cr
Sc II &2.08&0.12  & 13 & 	-1.09	&-0.12&&-0.86 &-0.24 &&-0.75 &-0.40 &&-1.37 &-0.21 \cr
Ti I &4.20&0.07   & 56 & 	-0.82	&+0.20&&-0.48 &+0.14 &&-1.20 &-0.20 &&-1.19 &-0.03 \cr
Ti II &4.22&0.14   & 8 & 	-0.80	&+0.17&&-     &-     &&-0.51 &-0.16 &&-1.16 & 0.00 \cr
V I &3.11&0.09    & 23 & 	-0.89	&+0.13&&-0.56 &+0.06 &&-1.24 &-0.24 &&-1.24 &-0.08 \cr
V II &3.01     &   & 1 & 	-0.99	&-0.02&&-     &-     &&-     & -    &&-     & -    \cr
Cr I &4.59&0.12   & 27 & 	-1.08	&-0.06&&-0.82 &-0.20 &&-1.19 &-0.19 &&-1.08 &+0.08 \cr
Cr II &4.85&0.08  & 6  & 	-0.82	&+0.15&&-     & -    &&-     & -    && -    &-     \cr
Mn I &4.24&0.10   & 11 & 	-1.15	&-0.13&&-1.00 &-0.38 &&-     & -    && -    &-     \cr
Fe I &6.48&0.07  & 180 & 	-1.02	& -   &&-0.62 & -    &&-1.00 & -    && -1.16&-     \cr
Fe II &6.53&0.07   & 17 & 	-0.97	& -  &&-     & -    &&-0.35 & -    && -1.16&-     \cr
Co I &4.11&0.12   & 21 & 	-0.81	&+0.21&&-0.70 &-0.08 &&-0.84 &+0.16 && -1.10&+0.06 \cr
Ni I &5.15&0.11   & 54 & 	-1.10	&-0.08&&-0.62 & 0.00 &&-0.84 &+0.16 && -1.12&+0.04 \cr
Cu I &3.31&0.17    & 3 & 	-0.90	&+0.12&&-     &-     &&-0.64 &+0.36 && -0.80&+0.36 \cr
Zn I &3.78    &    & 1 & 	-0.82	&+0.20&&-     &-     &&-0.39 &+0.61 && -0.92&+0.24 \cr
Y II &1.21&0.12    & 6 & 	-1.03	&-0.06&&-0.73 &-0.11 &&-0.62 &-0.27 && -1.14&+0.02 \cr
Zr I &     &       &   &        &     &&      &      &&-1.43 &-0.43 && -1.43&-0.27 \cr
Zr II &1.91&0.13   & 3 & 	-0.69	&+0.28&&-0.37 &+0.25 &&-     &-     && -    & -    \cr
Ba II &0.97&0.05   & 3 & 	-1.16	&-0.19&&-0.66 &-0.04 &&-     & -    &&-     &-     \cr
La II &0.24&0.25   & 3 & 	-0.93	&+0.04&&-0.52 &+0.10 &&-     &  -   &&-     &-     \cr

Ce II &0.26&0.16   & 5 & 	-1.32	&-0.35&&-0.65 &-0.03 &&-0.68 &-0.33 &&-1.21 &-0.05 \cr
Pr II &-0.20&0.13  & 2 & 	-0.91	&+0.06&&-     &-     &&-     & -    && -    & -    \cr
Nd II &0.56&0.13   & 8 & 	-0.94	&+0.03&&-     &-     &&-0.61 &-0.26 &&-1.15 &+0.01 \cr
Eu II &0.17 &      & 1 & 	-0.34	&+0.63&&-     &-     &&-0.40 &-0.05 &&-1.01 &+0.15 \cr
\hline
\end{tabular}
\end{table*}

\vspace{6pt}
\noindent{\it HD 139409}
\nopagebreak

\noindent The effective temperature of HD~139409 has been derived from the
excitation equilibrium  
of Fe\,{\sc i}, Ti\,{\sc i} and Cr\,{\sc i} lines 
(see Fig.~\ref{Fig:FEvsCHI} for Fe). 
The surface gravity $\log g$ was determined from the 
Fe\,{\sc i}/Fe\,{\sc ii} ionization balance, and the
microturbulent velocity by forcing the abundances of individual Fe\,{\sc i},
Ti\,{\sc i} and Cr\,{\sc i} lines to be independent of equivalent 
width (see Fig.~\ref{Fig:FEvsEW} for Fe). 
Although 
spectroscopic gravity and temperature determinations in late-type,
metal-deficient stars are probably affected by 
non-LTE effects,
these effects remain small when [Fe/H] $\ge -1.0$ \citep[see, for example,
][]{Allende-Prieto-1999}. The stellar parameters of Arcturus 
([Fe/H]$ = -0.6$, $\log g =
1.3$) derived by the spectroscopic method (applied on the HARPS
spectrum) are in good 
agreement indeed with those derived by other (non-spectroscopic) methods. 

The resulting atmospheric parameters for HD~139409 are as follows: 
$T_{\rm eff} = 5000$~K, $\log g = 2.8$ (cgs), and $\xi_t$ = 2.0~km~s$^{-1}$. 
The spectroscopic gravity, combined with a mass of 1~\Msun, leads to 
$M_V  \simeq +1.5$. Thus our
calculations indicate that HD~139409 is
less luminous than predicted by \citet{Mennessier-97}.
The derived iron abundance is [Fe/H] = $-0.42$ [adopting
$\log(\epsilon(\rm Fe)) = 7.50$].

The basic conclusion from the set of abundances listed in
Table~\ref{Tab:abundances139409} and displayed in Fig.~\ref{Fig:139409} is
that HD~139409 appears to be a mild barium
star.\footnote{Interestingly, \protect\citet{Pinsonneault-1984}
  quoting
a private communication from E. Luck, remark that Luck ``is completing
  an abundance analysis of HD~139409, which will
show that the light s-process elements do have ($\sim +1.5$~dex) enhancements,
while heavier s-process element enhancements are much smaller ($\sim
+0.3$~dex)''. Luck's study seems to have never been 
published.}  The s-process overabundances 
observed in HD~139409, although quite moderate, are not much smaller than
those observed  in a yellow symbiotic star like 
BD~-21$^\circ$3873, which exhibit overabundances of s-process elements in the range 
0.3 -- 0.8~dex \citep{Smith-97}.
With [Fe/H] = $-0.4$, HD~139409 is, however, not as
metal-deficient as the other stars considered in this paper.

\begin{table}

\caption[]{Abundances for
HD~139409, in the scale where $\log \epsilon$(H) = 12, and 
normalized with respect to the solar abundances \citep{Grevesse-1998}. 
The standard deviations $\sigma$ and the number $N$ of
lines used in the analysis are also given.}
\label{Tab:abundances139409}
\begin{tabular}{lrrrrr}
\hline\noalign{\smallskip}
  Element(X) & $Z$ & $\log \epsilon$(X) &  $\sigma$ &  $N$ & [X/Fe]  \\
\noalign{\smallskip}
\hline\noalign{\smallskip}
C\,{\sc i}   & 6 & 8.46   &  0.10     &  4 &  +0.36  \\
Na\,{\sc i}  &  11 & 6.02   &  0.05     &  5 &  +0.11  \\
Mg\,{\sc i}  &  12 & 7.31   &  0.12     &  6 &  +0.15  \\
Si\,{\sc i}  &  14 & 7.33   &  0.06     & 20 &  +0.20  \\
Ca\,{\sc i}  &  20 & 6.20   &  0.07     &  6 &  +0.26  \\
Sc\,{\sc ii} &  21 & 2.79   &  0.06     & 10 &  +0.04  \\
Ti\,{\sc i}  &  22 & 4.77   &  0.09     & 51 &  +0.17  \\
Ti\,{\sc ii} &  22 & 4.72   &  0.11     &  4 &  +0.12  \\
V\,{\sc i}   &  23 & 3.76   &  0.11     & 23 &  +0.18  \\
Cr\,{\sc i}  &  24 & 5.21   &  0.11     & 11 & $-$0.04 \\
Cr\,{\sc ii} &  24 & 5.36   &  0.09     &  3 &  +0.11  \\
Mn\,{\sc i}  &  25 & 5.13   &  0.11     &  9 &  +0.16  \\
Fe\,{\sc i}  &  26 & 7.08   &  0.08     &136 &   -  \\
Fe\,{\sc ii} &  26 & 7.08   &  0.12     & 14 &   -  \\
Ni\,{\sc i}  &  28 & 5.86   &  0.07     & 44 &  +0.03  \\
Y\,{\sc i}   &  39 & 2.22   &  0.10     &  2 &  +0.40  \\
Y\,{\sc ii}  &  39 & 2.41   &  0.07     &  6 &  +0.59  \\
Zr\,{\sc i}  &  40 & 2.58   &  0.09     &  3 &  +0.40  \\
Zr\,{\sc ii} &  40 & 2.78   &  0.12     &  3 &  +0.60  \\
Ba\,{\sc ii} &  56 & 1.89   &  0.12     &  3 &  +0.18  \\
La\,{\sc ii} &  57 & 1.05   &  0.12     &  3 &  +0.30  \\
Ce\,{\sc ii} &  58 & 1.46   &  0.14     &  5 &  +0.30  \\
Nd\,{\sc ii} &  60 & 1.38   &  0.09     &  6 &  +0.30  \\
\noalign{\smallskip}
\hline
\end{tabular}
\end{table}

\section{Statistics of binarity among metal-deficient Ba stars}
\label{Sect:Binary}

The list of confirmed or suspected metal-deficient barium stars remaining after the
screening process based on abundances (as described in 
Sect.~\ref{Sect:abundances}) is given in Table~\ref{Tab:SB}.
The possible binary nature of those stars may be assessed using three
different methods:
\begin{itemize}
\item checking for radial-velocity variations;
\item checking for astrometric orbital motion, directly from the Hipparcos
astrometric data;
\item checking for astrometric orbital motion, indirectly from a comparison 
of the Hipparcos and Tycho-2 proper motions.
\end{itemize}

These methods are now described in turn.

\begin{table*}
\caption[]{\label{Tab:SB}
Binary properties of confirmed or suspected metal-deficient barium stars. The
column labeled `Ba' indicates whether detailed chemical analyses have
confirmed the Ba nature of the star. The columns labeled 
 $\chi^2_{\mu_{\rm HIP} - \mu_{\rm Tyc}}$	and Prob($\chi^2 >
\chi^2_{\mu_{\rm HIP} - \mu_{\rm Tyc}})$
provide the $\chi^2$  (and its associated probability) involved in the comparison of Hipparcos and Tycho-2 proper motions
(see text for details). The column 'Bin.' has been set to 'y' if the first kind risk of rejecting the null hypothesis
that the Tycho and Hipparcos proper motions are equal is smaller than 10\%. The column labeled IAD
indicates whether the signature of an orbital motion is present in the Hipparcos Intermediate Astrometric Data, according
to the  various tests described in the text. The column labeled $\sigma(Vr)$  provides the radial-velocity standard
deviation,
$\Delta t$ and $N$ correspond to the time span and number of  observations, respectively. The columns labeled
$P$ and `Ref' list the orbital period (when available) and the reference for the radial velocity and/or orbital data. The
column 'Binary' gives the final binarity diagnostics.  }
\tabcolsep 4pt
\begin{tabular}{rrccrcccclllccc}
\hline
 HIP  &  HD/DM  & Ba & Ref & \multicolumn{1}{c}{$\varpi$ (mas)} &
 \multicolumn{3}{c}{$\mu_{\rm HIP} - \mu_{\rm Tyc}$} & IAD &
$\sigma(Vr)$ & 
$\Delta t$ & $N$ &
$P$  & Ref & Binary  \cr
\cline{6-8}
      &         &     &    &                 & $\chi^2_{\rm obs}$ & Prob($\chi^2 > \chi^2_{\rm obs})$ & Bin. & &    (km
s$^{-1}$) & (d) & & (d)\cr
\hline
  4347 & 5424  & & & $0.22\pm1.42$ & 1.05 & 0.59 & n & n  & 2.32 & 3306 & 13 & 1881 & 6 & y \cr
 11595 & 15589 & & & $2.03\pm1.21$ & 1.63 & 0.44 & n & n  & -    &  -   & -  &  -   &   & ? \cr
 25161 & $-27^\circ$2233 & & & $0.89\pm1.35$ & 0.23 & 0.89 & n & n & -   & -  &  - & -   &   & ? \cr
 29740 &43389  & & & $-1.25\pm1.00$  & 4.54 & 0.10 & y & y & 3.85 &3350& 24 & 1689 & 6 & y \cr
 34795 &55496  & y & 5 & $2.44\pm1.04$  & 0.02 & 0.99 & n & y & 0.73 & 5121 & 24&- & 2 & y? \cr
58596 &104340  & y & 3 & $-0.97\pm1.09$ & 3.39 & 0.18 & n & n & 1.64 & 2587 &16 & - & 2 & n  \cr
 69834 &123396 &   &   & $1.73\pm0.86$ & 1.98 & 0.27 & n & n & -  & -   & -  & -    &   & ? \cr
 76605 &139409 & y & 2 & $5.51\pm1.14$ & 4.98 & 0.08 & y & n & 0.66 & 1478 & 2 & -  & 2 & y?\cr
 97874 &187762 &   &   & $2.07\pm1.53$ & 9.00 & 0.01 & y & y & -  & -  &  - &  -   &   & y \cr 
107478 &206983 & y & 3 & $3.75\pm1.86$ & 2.94 & 0.23 & n & y & -  & -  &  - &  -   &   & y?\\
\hline
\medskip\\
\noalign{Additional stars [from refs. (1), (5) and (10)]}
\medskip\\
\hline
43042 &+$75^\circ 348$ &  y & 1 & $1.02\pm1.32$ & 0.12 & 0.94  & n & y & 4.64 & 1436 & 33 & 1042 & 11 & y  \cr 
55852 & +$4^\circ 2466$ &  y & 4,5 & $0.96\pm1.83$ & 0.43 & 0.81 & n & n & 3.38 & 6191 & 41 & 4592 & 2  & y \cr
       & +$3^\circ 2688$ &  y & 5 & - & -  & -& -& - &0.45 & 986 & 7 & - & 2 & n? \cr
      &CS 22942-019  & y & 10 & - & -  &- & -& - & 3.42 & 3274 & 15 & 2800 & 10 & y \cr
      &CS 22948-027  & y & 10 & - & -  &- &- & - & 1.87 & 2560 & 13 & 505  & 10 & y \cr
\hline
\end{tabular}
References to the Table:\\
(1) \citet{Zacs-2000} (2) This work (3) \citet{Junqueira-2001}
(4) \citet{Burris-2000} (5) \citet{Luck-Bond-91} (6)
\citet{Udry-1998:a} 
(10) \citet{Preston-Sneden-01}  
(11) \cite{Zacs-2005} 
\end{table*}

\subsection{Radial-velocity variations}

Several stars from Table~\ref{Tab:SB} (namely HIP~4347, HIP~29740, HIP~34795, HIP~55852, HIP~58956, HIP~76605 and 
BD +3$^\circ2688$)
have been monitored for many years using the  CORAVEL spectrovelocimeter
\citep{Baranne-1979}, as part of a larger program aiming at finding the
frequency of spectroscopic binaries among s-process-rich late-type giants
\citep[see ][for details and other results from this CORAVEL
monitoring]{Jorissen-Mayor-88,Jorissen-Mayor-92,Jorissen-VE-98}. Individual radial-velocity measurements for those stars
\citep[in the CORAVEL-ELODIE system as defined in ][]{Udry-1999} are given in Tables~\ref{Tab:RV} or in \citet{Udry-1998:a}.
For a few other stars (HIP~43042, CS~22942-019 and CS~22948-027), radial velocities  were monitored using other
instruments, and their results were taken from the literature \citep{Preston-Sneden-01, Zacs-2005}.

\begin{figure}
\resizebox{\hsize}{!}{\includegraphics{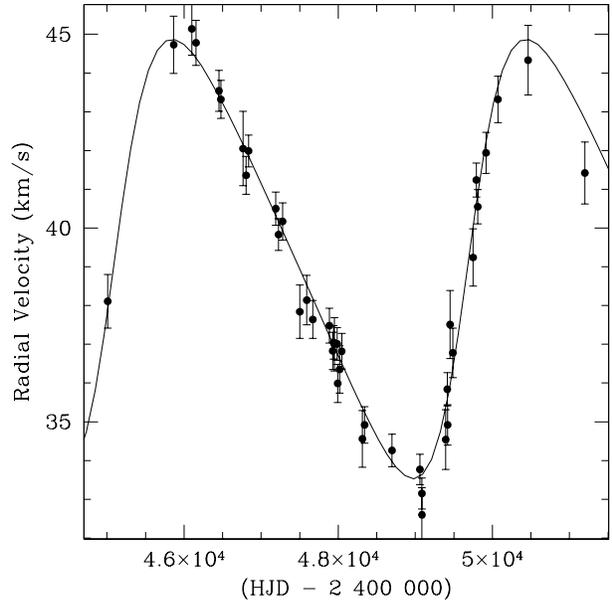}}
\caption[]{\label{Fig:bd04_2366}
Radial velocities as a function of heliocentric Julian Day for BD +4$^\circ$2466, 
superimposed on the orbital solution corresponding to the orbital elements listed
in Table~\protect\ref{Tab:SB4_2466}. }
\end{figure}

Orbits were already available for HIP~4347 and HIP~29740
\citep{Udry-1998:a}, HIP~43042 \citep{Zacs-2005}, as well as for CS~22942-019 and  CS~22948-027
\citep{Preston-Sneden-01}.  A new orbital solution has been derived for
BD~+04$^\circ$2466 (Table~\ref{Tab:SB4_2466} and Fig.~\ref{Fig:bd04_2366}). 
The binary nature of those stars is therefore beyond doubt.

\begin{figure}
\resizebox{\hsize}{!}{\includegraphics{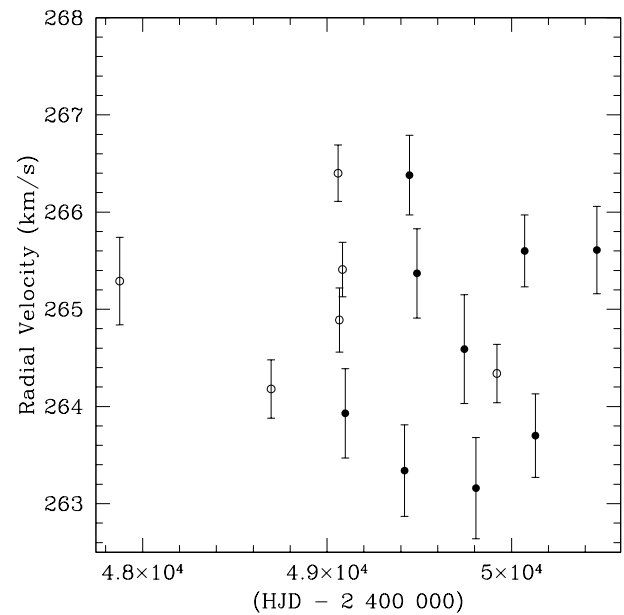}}
\caption[]{\label{Fig:104340_RV}
Radial velocities as a function of heliocentric Julian Day for
HIP~58596 = HD~104340 (open circles correspond to measurements obtained
with the southern CORAVEL, and filled circles with the northern one).  
}
\end{figure}

The radial-velocity standard deviation  of  HIP~58596 is larger than expected based on the
uncertainty on one measurement (Fig.~\ref{Fig:104340_RV}), but no satisfactory orbital solution could be found. 
A 3-d orbit (with an eccentricity of 0.30) is possible, but this short
orbital period is not consistent with the giant nature of HIP~58596,
which imposes orbital periods of at least 20~d \citep[see Fig.~4 of ][]{Pourbaix-04a}.
The large standard deviation exhibited by HIP~58596 is therefore very likely another example of the large intrinsic jitter
often observed for  metal-deficient stars, as discussed by  \citet{McClure-84a} and \citet{Carney-2003}. Nevertheless, the proper motion analysis presented in Sect.~\ref{Sect:Tycho} hints at a possible
long orbital period  for HIP~58596.

A 14~y radial-velocity monitoring for HIP~34795 with the northern and southern CORAVELs 
is not very conclusive either, mostly because there are difficulties
in finding the zero-point offset between the two instruments for such large radial velocities \citep{Udry-1999}.
When a $-1$~\kms\ offset is applied to the northern velocities (with
respect to the values listed in Table~\ref{Tab:RV}), a long-term trend
seems to be present, albeit with some superimposed jitter (Fig.~\ref{Fig:55496_RV}). The analysis of the Hipparcos
astrometric data presented in Sect.~\ref{Sect:IAD} suggests that the star might be binary, although the evidence is not
very conclusive. 

\begin{figure}
\resizebox{\hsize}{!}{\includegraphics{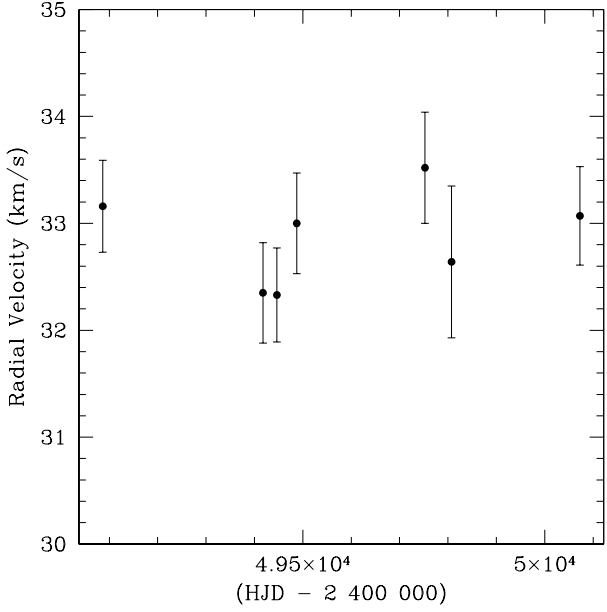}}
\caption[]{\label{Fig:bd03_2688}
Radial velocities as a function of heliocentric Julian Day for BD +3$^\circ$2688.
}
\end{figure}

\begin{figure}
\resizebox{\hsize}{!}{\includegraphics{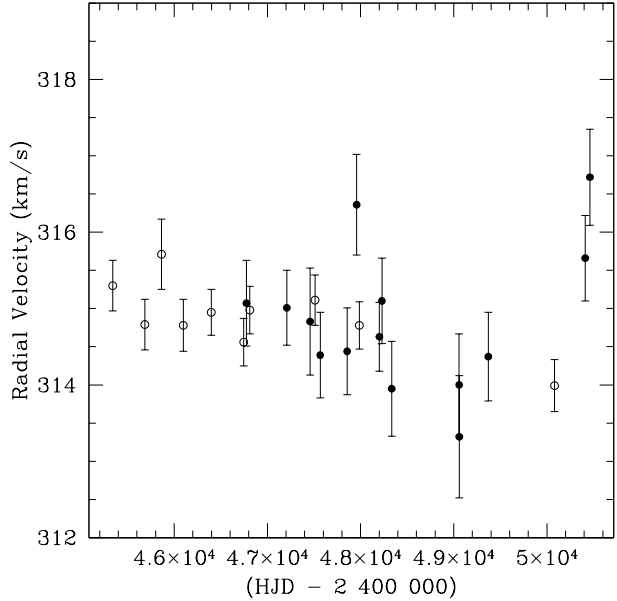}}

\caption[]{\label{Fig:55496_RV}
Radial velocities as a function of heliocentric Julian Day for
HIP~34795 = HD~55496. The symbols are as in Fig.~\protect\ref{Fig:104340_RV}.
With respect to the data listed in Table~\protect\ref{Tab:RV}, 
an offset of $-1$~\kms\ has been applied to the northern measurements to make them
consistent with the southern ones.  
 }
\end{figure}

Finally, there is no sign of radial-velocity variations for BD~+$3^\circ2688$
(Fig.~\ref{Fig:bd03_2688}).

\begin{table}
\caption[]{\label{Tab:SB4_2466}
Orbital elements for BD +04$^\circ$2466.
}
\begin{tabular}{lllllll}
\hline
$P$ (d)           &4592.7$\pm 51.1$ \cr
$e$               &0.286$\pm0.02$ \cr 
$\omega$ (deg)    &266.8$\pm5.3$ \cr
$V_\gamma$ (\kms) &+39.29$\pm0.11$\cr 
$K$ (\kms)        &5.67$\pm0.12$ \cr 
$T$ (JD)          & 2\ts445\ts076.6$\pm 73.0 $\cr 
$f(M)$ (\Msun)    & 0.076\cr
$a_1 \sin i$ (Gm) & 343.18\cr
$N$               & 41 \cr
\hline
\end{tabular}
\end{table} 

\renewcommand{\baselinestretch}{1}
\begin{table}[t]
\caption[]{\label{Tab:RV}
Individual CORAVEL radial velocities and associated errors for HIP~34795, HIP~58596, HIP~76605,
BD~+$3^\circ2688$ and BD~+04$^\circ$2466. The last column indicates which 
one of the two CORAVEL spectrovelocimeters has been used (NO = CORAVEL north
at the {\it Observatoire de Haute Provence}; SO = CORAVEL south on the Danish 1.54-m telescope
at ESO)\\
}
\begin{tabular}{ccccc}
\multicolumn{4}{c}{\bf HIP 34795 = HD 55496}\cr
\hline
{\small DDMMYY} & HJD & R.V.   & $\epsilon$ \cr
                &     & (\kms) & (\kms) \cr
\hline 
050183& 45340.695 & 315.30 &0.33 &SO\\
151283& 45684.797 & 314.79 &0.33 &SO\\
120684& 45864.434 & 315.71 &0.46 &SO\\
310185& 46097.658 & 314.78 &0.34 &SO\\
271185& 46397.826 & 314.95 &0.30 &SO\\
111186& 46746.858 & 314.56 &0.31 &SO\\
101286& 46775.595 & 316.07 &0.56 &NO\\
150187& 46811.682 & 314.98 &0.31 &SO\\
160288& 47208.394 & 316.01 &0.49 &NO\\
231088& 47458.704 & 315.83 &0.70 &NO\\
151288& 47511.857 & 315.11 &0.33 &SO\\
090289& 47567.393 & 315.39 &0.56 &NO\\
241189& 47855.642 & 315.44 &0.57 &NO\\
070390& 47958.339 & 317.36 &0.66 &NO\\
050490& 47987.497 & 314.78 &0.31 &SO\\
061190& 48202.697 & 315.63 &0.45 &NO\\
031290& 48229.645 & 316.10 &0.56 &NO\\
180391& 48334.380 & 314.95 &0.62 &NO\\
110393& 49058.361 & 315.00 &0.67 &NO\\
130393& 49060.338 & 314.32 &0.80 &NO\\
180194& 49371.471 & 315.37 &0.58 &NO\\
301295& 50082.752 & 313.99 &0.34 &SO\\ 
231196& 50411.600&  316.66& 0.56& NO\\
120197& 50461.480&  317.72& 0.63& NO\\ 
\hline
\medskip\\
\multicolumn{4}{c}{\bf HIP 58596 = HD 104340}\cr
\hline
{\small DDMMYY} & HJD & R.V.   & $\epsilon$ \cr
                &     & (\kms) & (\kms) \cr
\hline 
 141289& 47875.832&  265.29& 0.45 & SO\cr
 140392& 48696.805&  264.18& 0.30 & SO\cr
 120393& 49059.755&  266.40& 0.29 & SO\cr
 190393& 49066.707&  264.89& 0.33 & SO\cr
 050493& 49083.708&  265.41& 0.28 & SO\cr
 200493& 49098.453&  263.93& 0.46 & NO\cr
 080394& 49420.525&  263.34& 0.47 & NO\cr
 030494& 49446.435&  266.38& 0.41 & NO\cr
 140594& 49487.360&  265.37& 0.46 & NO\cr
 260195& 49744.638&  264.59& 0.56 & NO\cr
 300395& 49807.475&  263.16& 0.52 & NO\cr
 210795& 49920.516&  264.34& 0.30 & SO\cr
 191295& 50071.721&  265.60& 0.37 & NO\cr
 150296& 50129.556&  263.70& 0.43 & NO\cr
 130197& 50462.684&  265.61& 0.45 & NO\cr
\hline
\medskip\\
\multicolumn{4}{c}{\bf HIP 76605 = HD 139409}\cr
\hline
{\small DDMMYY} & HJD & R.V.   & $\epsilon$ \cr
                &     & (\kms) & (\kms) \cr
\hline 
 090790 &48082.629 &  62.43 &0.30 & SO \cr
 260794 &49560.513 &  63.36 &0.27 & SO \cr
\hline
\end{tabular}
\end{table}

\addtocounter{table}{-1}
\begin{table}
\caption[]{Continued. \\
}
\begin{tabular}{ccccc}
\multicolumn{4}{c}{\bf HIP 55852 = BD +04$^\circ$2466}\cr
\hline
{\small DDMMYY} & HJD & R.V.   & $\epsilon$ \cr
                &     & (\kms) & (\kms) \cr
\hline 
070282&  45008.814&   38.11& 0.69 & SO\cr
130684&  45865.491&   44.73& 0.74 & SO\cr
030285&  46100.832&   45.14& 0.68 & SO\cr
290385&  46154.675&   44.78& 0.58 & SO\cr
200186&  46451.831&   43.54& 0.53 & SO\cr
140286&  46476.783&   43.32& 0.49 & SO\cr
291186&  46764.706&   42.05& 0.96 & NO\cr
090187&  46805.864&   41.36& 0.49 & SO\cr
080287&  46835.808&   41.99& 0.41 & SO\cr
280188&  47189.812&   40.50& 0.43 & SO\cr
040388&  47225.726&   39.83& 0.41 & SO\cr
250488&  47277.657&   40.17& 0.48 & SO\cr
051288&  47501.707&   37.84& 0.69 & NO\cr
060389&  47592.565&   38.14& 0.64 & NO\cr
220589&  47669.526&   37.64& 0.49 & SO\cr
241289&  47885.857&   37.48& 0.45 & SO\cr
050290&  47928.800&   36.83& 0.48 & SO\cr
160290&  47939.743&   37.05& 0.43 & SO\cr
260290&  47949.595&   37.00& 0.69 & NO\cr
040490&  47986.631&   37.01& 0.42 & SO\cr
110490&  47993.594&   35.99& 0.49 & SO\cr
050590&  48017.647&   36.35& 0.61 & SO\cr
020690&  48045.501&   36.82& 0.46 & SO\cr
260291&  48314.487&   34.56& 0.73 & NO\cr
260391&  48342.724&   34.92& 0.47 & SO\cr
150392&  48697.747&   34.26& 0.42 & SO\cr
120393&  49059.726&   33.77& 0.39 & SO\cr
070493&  49085.670&   33.15& 0.40 & SO\cr
080493&  49086.448&   32.60& 0.70 & NO\cr
100294&  49394.609&   34.54& 0.77 & NO\cr
020394&  49414.713&   35.84& 0.43 & SO\cr
070394&  49419.515&   34.92& 0.52 & NO\cr
080494&  49451.424&   37.51& 0.88 & NO\cr
140594&  49487.351&   36.78& 0.64 & NO\cr
310195&  49749.592&   39.24& 0.74 & NO\cr
150395&  49792.681&   41.24& 0.44 & SO\cr
020495&  49810.460&   40.55& 0.44 & NO\cr
190795&  49918.506&   41.94& 0.53 & SO\cr
191295&  50071.715&   43.32& 0.60 & NO\cr
130197&  50462.674&   44.33& 0.90 & NO\cr
200199&  51199.634&   41.42& 0.80 & NO\cr
\hline
\medskip\\
\multicolumn{4}{c}{\bf BD +03$^\circ$2688}\cr
\hline
{\small DDMMYY} & HJD & R.V.   & $\epsilon$ \cr
                &     & (\kms) & (\kms) \cr
\hline 
 080493 &49086.481 &  33.16 &0.43 & NO \cr
 050394 &49417.506 &  32.35 &0.47 & NO \cr
 030494 &49446.510 &  32.33 &0.44 & NO \cr
 140594 &49487.388 &  33.00 &0.47 & NO \cr
 030295 &49752.658 &  33.52 &0.52 & NO \cr
 300395 &49807.506 &  32.64 &0.71 & NO \cr
 201295 &50072.729 &  33.07 &0.46 & NO \cr
\hline
\end{tabular}
\end{table}

\subsection{Orbital motion from Hipparcos astrometric data}
\label{Sect:IAD}

A tailored reprocessing of the Hipparcos  {\it Intermediate Astrometric Data}
\citep[hereafter IAD; ][]{vanLeeuwen-1998:a} makes it
possible to look for a possible orbital signature in the astrometric motion,
following the method outlined by
\citet{Pourbaix-2000:b}, \citet{Pourbaix-Boffin:2003},
\citet{Pourbaix-2004} and applied to barium stars by
\citet{Jorissen-2004}. We give here only a brief summary of the method.

The basic idea is to quantify the likelihood of the fit of the Hipparcos
astrometric data with an orbital model. For that purpose,
\citet{Pourbaix-2001:b} \citep[see also ][]{Jancart-2005}
introduced several  statistical indicators which 
allow us to decide whether to keep or to discard an orbital solution.  Those
indicators relevant to our purpose are the following:
\begin{itemize}
\item The addition of 4 supplementary parameters (the four Thiele-Innes orbital 
constants) describing the orbital motion should result in a statistically
significant decrease of the $\chi^2$ for the fit of the $N$ IAD with an orbital
model with 9 free parameters ($\chi^2_T$), 
as compared to a fit with a single-star solution with 5 free parameters ($\chi^2_S$).
This criterion is expressed by an $F$-test:
\begin{equation}
Pr_2 = Pr[\hat{F} > F(4,N-9)],
\end{equation}
where

\begin{equation} 
\hat{F} =  
\frac{N-9}{4}\;\; \frac{\chi^2_S - \chi^2_T}{\chi^2_T}.
\end{equation}
$Pr_2$ is thus the first kind risk associated with the rejection of the null 
hypothesis:``{\em there is no orbital wobble present in the data}''.

\item Getting a substantial reduction of the $\chi^2$ with the Thiele-Innes model 
does not necessarily   imply that the four Thiele-Innes constants $A,B,F,G$ are
significantly different from 0. The first kind risk associated with the rejection
of the null hypothesis ``{\em the orbital semi-major axis is equal to zero}'' may
be expressed as
\begin{equation}
Pr_3 = Pr[\chi^2_{ABFG} > \chi^2(4)],
\end{equation}
where
\begin{equation}
\chi^2_{ABFG} =  
\vec{X}^t \vec{V}^{-1} \vec{X},
\end{equation}
and $\vec{X}$ is the vector of components $A,B,F,G$ and $\vec{V}$ is its 
covariance matrix.\footnote{Since it may be shown that 
$\chi^2_{S} -  \chi^2_{T} = \chi^2_{ABFG}$, the $Pr_2$ and $Pr_3$ tests are in fact equivalent 
provided that $\chi^2_{T} \sim N-9$. Thus, if $Pr_2$ and $Pr_3$ are significantly different, it 
means either that the Thiele-Innes orbital model does not fit the data very well 
($\chi^2_{T} >> N-9$), or that it fits much better than could be expected ($\chi^2_{T} << N-9$).
We are indebted to L. Lindegren for this clarification 
\protect\citep[see also ][]{Jancart-2005}.}

\item For the orbital solution to be   a significant one, its parameters should
not be strongly correlated with the  other astrometric parameters
(e.g., the proper motion). In other words, the covariance matrix of
the astrometric solution should be dominated by its diagonal terms, as
measured by the {\it efficiency}
$\epsilon$ of the matrix being close to 1 \citep{Eichhorn-1989}. The efficiency is simply expressed by 
\begin{equation}
\epsilon = \sqrt[m]{\frac{\Pi_{k=1}^m \lambda_k}{\Pi_{k=1}^m \vec{V}_{kk}}},
\end{equation}
where $\lambda_k$ and $\vec{V}_{kk}$ are respectively the eigenvalues and the 
diagonal terms of the covariance matrix $\vec{V}$.
\end{itemize}

With the above  notations, the requirements for a star to
qualify as a binary is then
\begin{equation}
\label{Eq:alpha}
\alpha \equiv (Pr_2 + Pr_3)/\epsilon \le 0.02,
\end{equation}
where the threshold value of 0.02 has been chosen to minimize false
detections
\citep{Jorissen-2004}.

Hipparcos data are, however, seldom precise enough to derive
the orbital elements from scratch. Therefore, when a spectroscopic orbit is
available beforehand, it is advantageous to import $e, P, T$ from the
spectroscopic orbit and to derive the remaining astrometric elements
\citep[as done by ][]{Pourbaix-2000:b,Pourbaix-Boffin:2003}.  If a spectroscopic
orbit is not available, trial $(e, P, T)$ triplets scanning a regular grid (with
$10 \le P (\rm d) \le 5000$ imposed by the Hipparcos scanning law and the mission
duration) may be used. The quality factor
$\alpha$ is then computed for each trial $(e, P, T)$ triplet, and if
there exist triplets yielding $\alpha < 0.02$, 
the star is flagged as a binary.
To test its success rate, this method has been applied by \citet{Jorissen-2004} on
a sample of barium stars. 
These authors show that, when  
$\varpi > 5$~mas and  $100 < P (\rm d)< 4000$, 
the (astrometric) binary detection rate is close to 100\%, {\it i.e.}, the astrometric
method recovers all known spectroscopic binaries \citep[see also ][]{Jancart-2005}.
When the orbit is not known beforehand, the method makes it even
possible to find a good estimate for the orbital period, provided, however, that
the true period is not an integer fraction, or a multiple, of one year.
Here the method is applied to the sample of metal-deficient barium stars
listed in Table~\ref{Tab:SB}.

\begin{figure*}
\resizebox{0.49 \hsize}{!}{\includegraphics{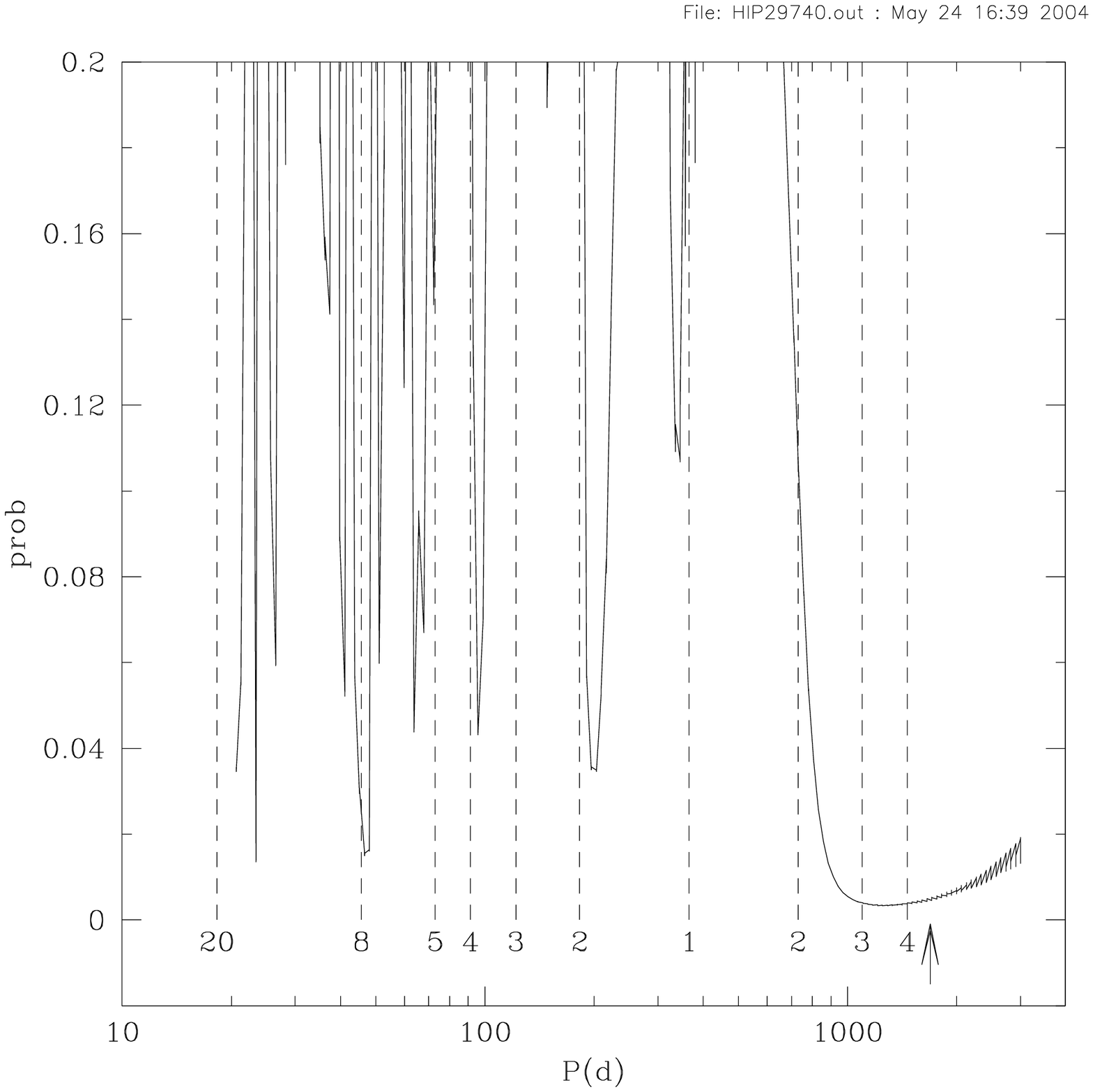}}
\resizebox{0.49 \hsize}{!}{\includegraphics{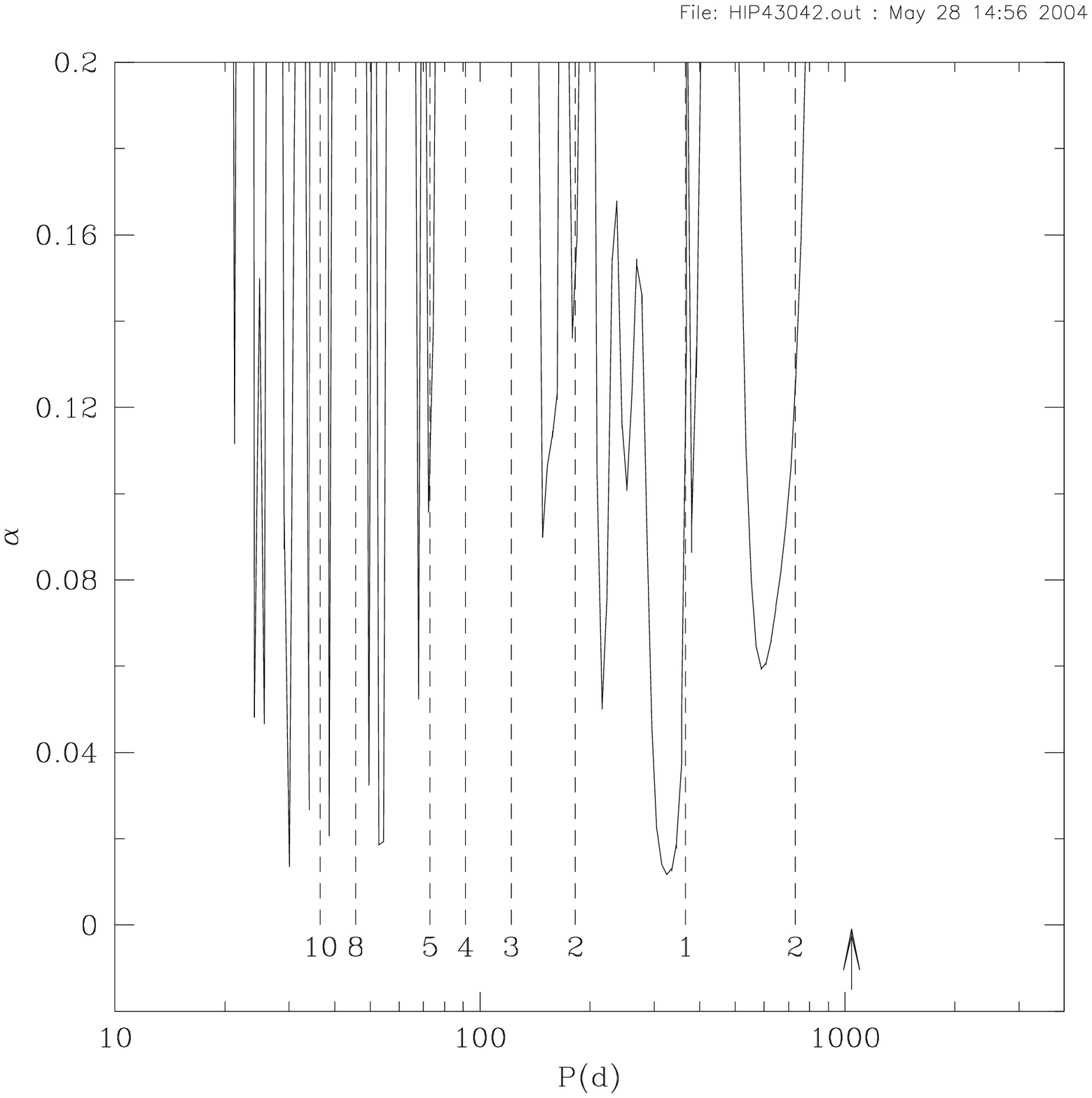}}\\
\resizebox{0.49 \hsize}{!}{\includegraphics{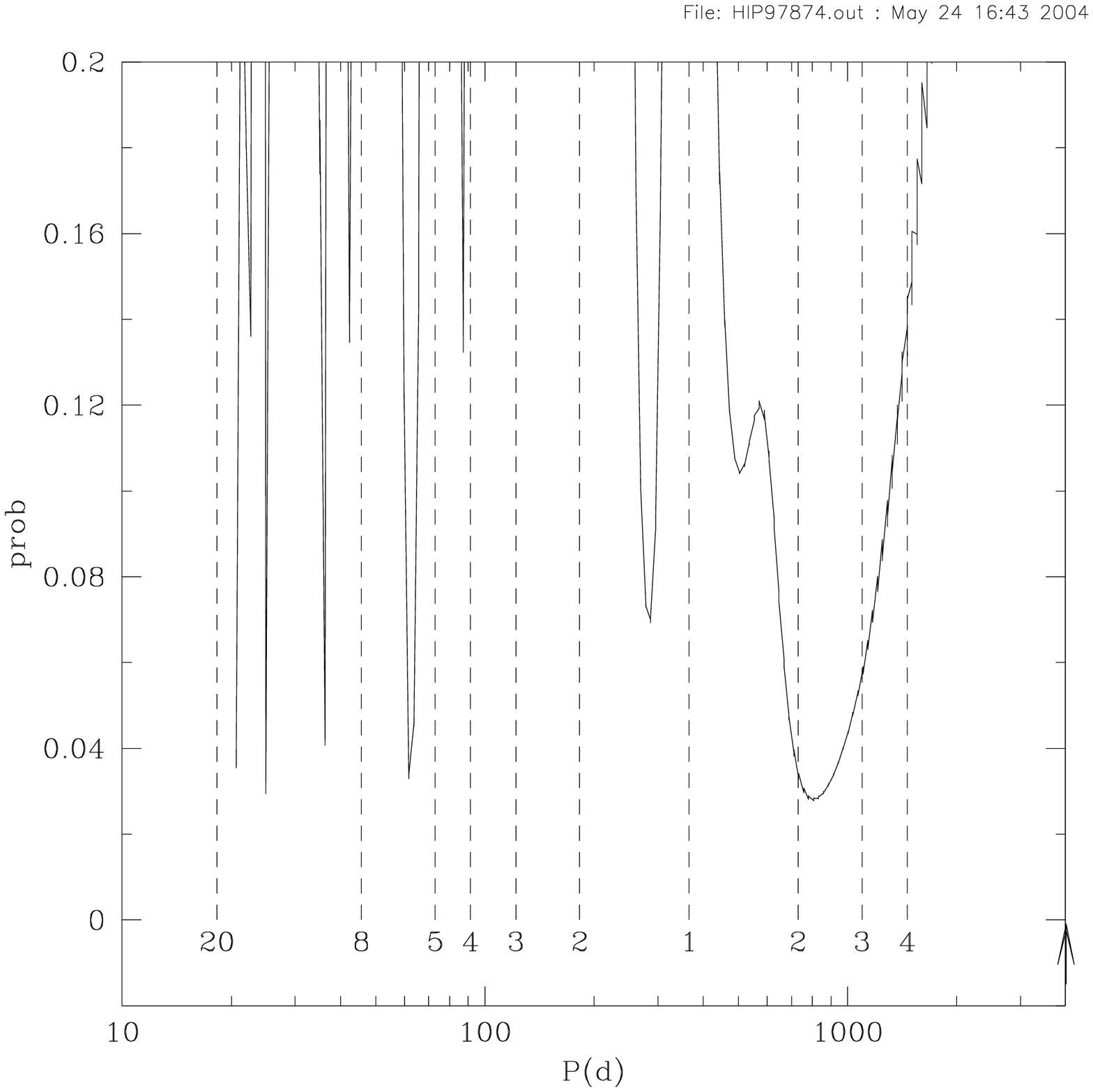}}
\resizebox{0.49 \hsize}{!}{\includegraphics{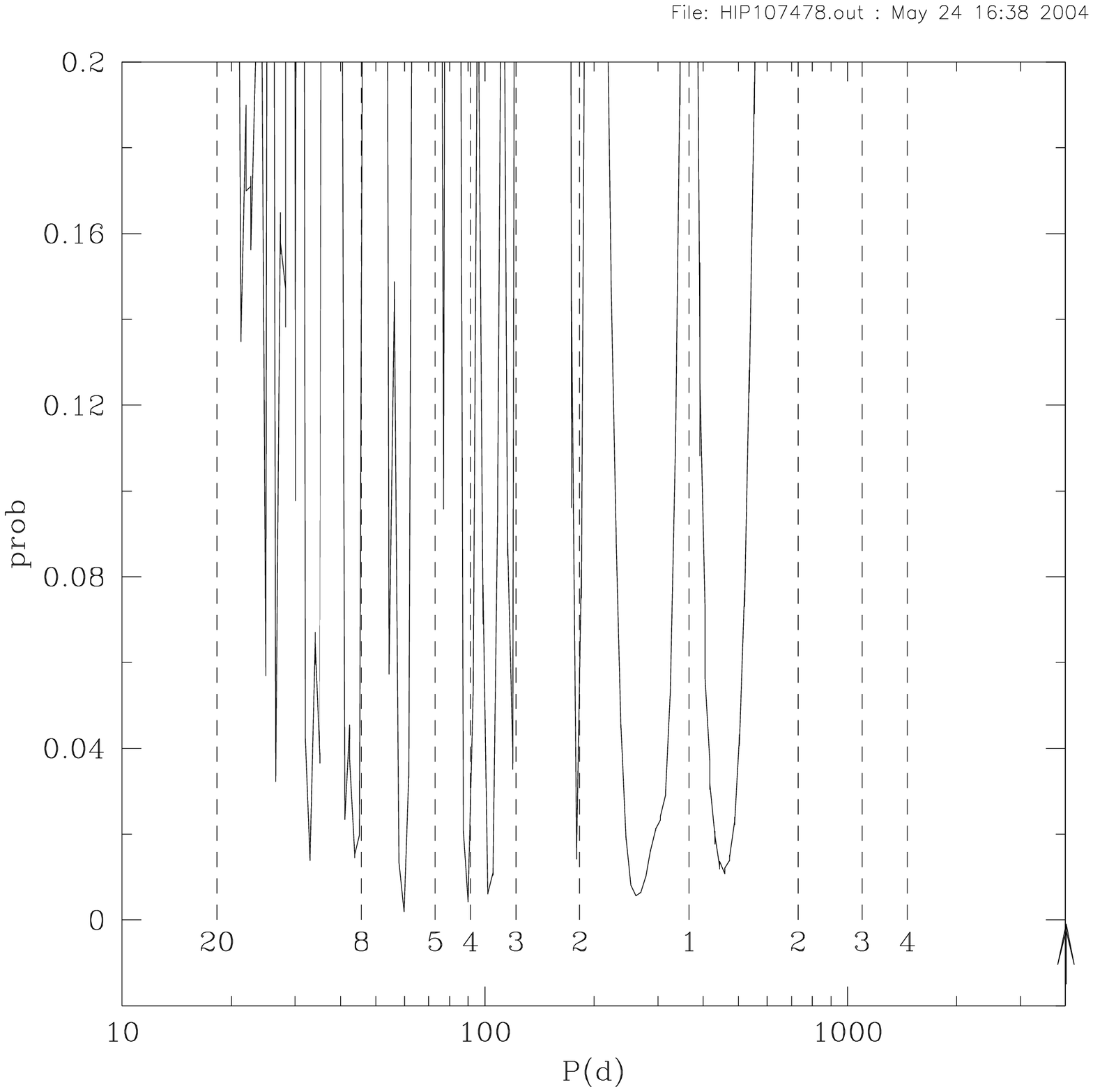}}
\caption{\label{Fig:alpha1}
The $\alpha$ statistics (Eq.~\protect\ref{Eq:alpha}) 
as a function of the trial orbital period (assuming $e
= 0$) for metal-deficient barium stars. 
The period from the spectroscopic orbit (represented by an arrow) of HIP~29740 (upper left panel)
indeed lies within the range of minimum 
$\alpha$ values. By comparison, the suspected binaries 
HIP~97874 and HIP~107478 are likely to have periods $P \sim 800$~d and 
$\sim 250-450$~d,  respectively. For comparison, stars non-flagged as
binaries have no 
$(e, P, T)$ grid points with $\alpha < 0.02$ (see
Fig.~\protect\ref{Fig:alpha2}).  
The vertical dashed lines represent multiple, or integer fractions, of
1~y. At those periods, there 
is a strong correlation between the parallactic and orbital signals, which degrades the $\alpha$ statistics
and makes binaries difficult to find at those 1-y alias periods.
} 
\end{figure*}

\begin{figure*}
\resizebox{0.49 \hsize}{!}{\includegraphics{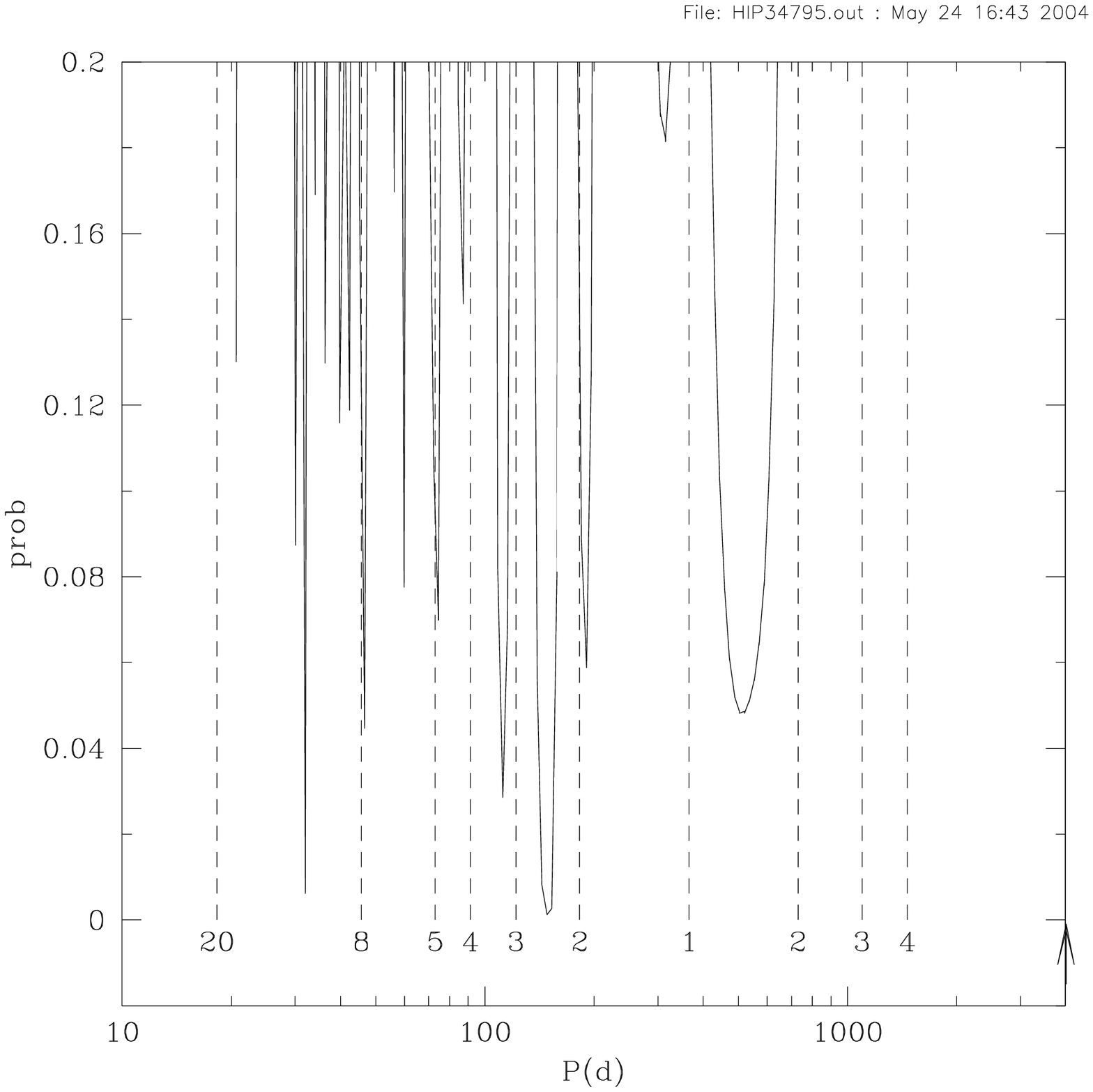}}
\resizebox{0.49 \hsize}{!}{\includegraphics{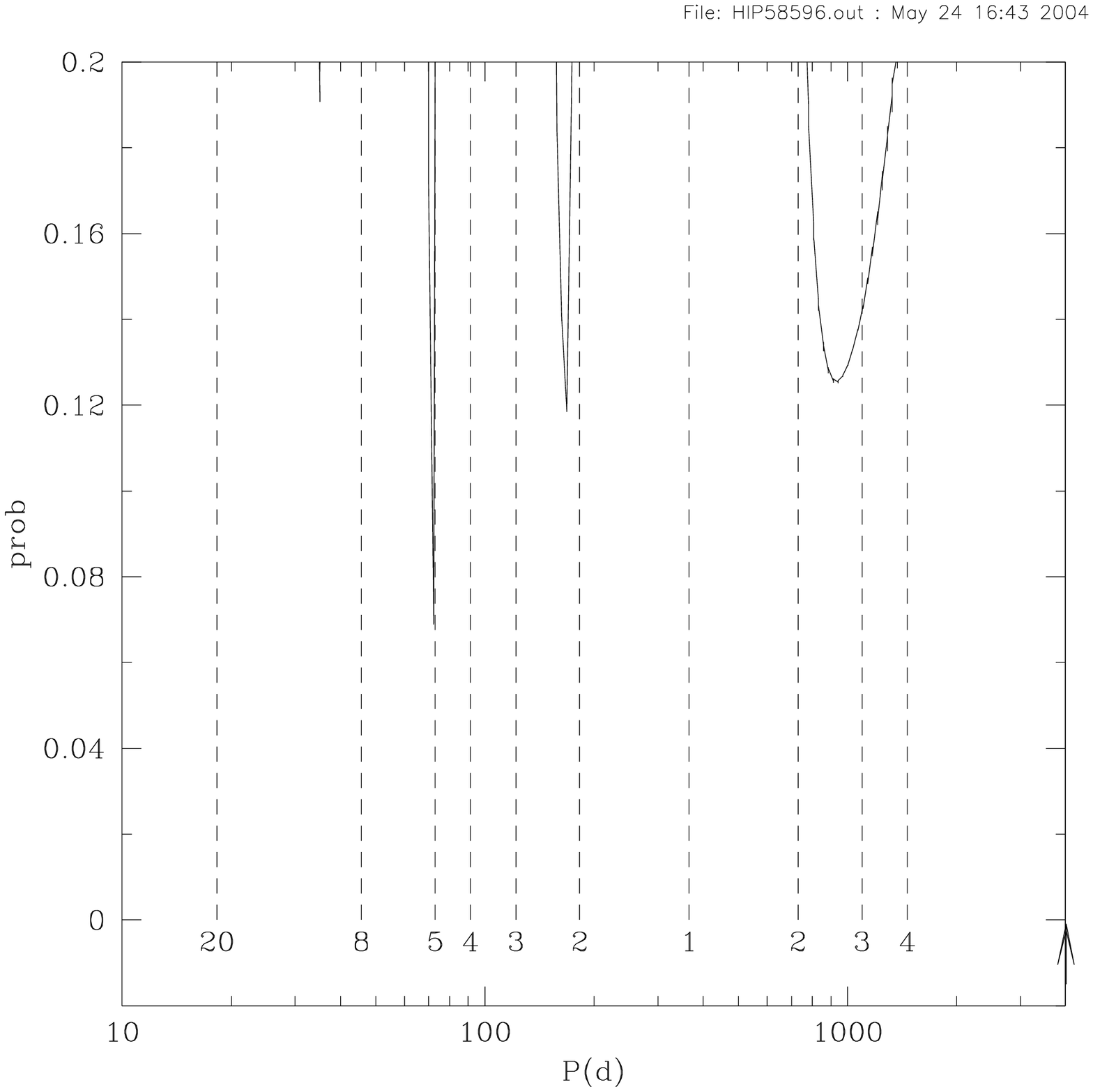}}
\caption{\label{Fig:alpha2}
Same as Fig.~\protect\ref{Fig:alpha1} for HIP~34795 (left panel,
flagged as binary) and HIP~58596 (right panel, non-binary).
} 
\end{figure*}

The method flags as definite binaries the stars HIP~29740, 34795,
43042, 97874 and 107478 (Figs.~\ref{Fig:alpha1} and
\ref{Fig:alpha2}). 
In two cases (HIP~29740 and 43042), 
the IAD method thus confirms the conclusion from the radial-velocity
monitoring, but yields as well three new binaries
(HIP~34795, 97874 and 107478). Two spectroscopic binaries (HIP~4347 and HIP~55852) are not detected by 
the IAD method because of their small
parallax or long orbital period.  The non-binary nature of
HIP~58596, already suspected from the radial-velocity data, is confirmed by the analysis of 
the IAD (Fig.~\ref{Fig:alpha2}).

\subsection{Orbital motion from a comparison of Hipparcos and Tycho-2  proper
motions} 
\label{Sect:Tycho}

\citet{Kaplan-Makarov-03}   suggested that the comparison of 
Hipparcos and Tycho-2 \citep{Hog-00} proper motions offers a way to detect binaries with long periods
(typically from 2000 to 4000~d).
The Hipparcos proper motion,
being based on observations  spanning only 3~y, may be altered by the orbital
motion, especially for systems with periods in the range of 2000 to 4000~d
whose orbital motion was not recognized by Hipparcos. On the other hand,
this effect should average out in the Tycho-2 proper motion, which is
derived from observations covering a much longer time span. 
This method, already used by \citet{Makarov-2004}, \citet{Pourbaix-2004} and \citet{Jancart-2006},  
works best when applied to stars with parallaxes in excess of about 5~mas.

The method evaluates the quantity
\begin{equation}
\chi^2_{\rm obs} = (\vec{\mu}_{\rm HIP} - \vec{\mu}_{\rm Tyc})^t\;  {\bf W}^{-1}\;  (\vec{\mu}_{\rm HIP} - \vec{\mu}_{\rm Tyc}),    
\end{equation}
where $\vec{\mu}_{\rm HIP}$ and $\vec{\mu}_{\rm Tyc}$ are the vectors of $\alpha$ and $\delta$ components of the Hipparcos and
Tycho-2 proper motions, respectively, and  {\bf W} is the associated $2\times2$ variance-covariance matrix. The
covariance between
$\mu_{\alpha,{\rm HIP}}$ and $\mu_{\delta,{\rm HIP}}$, as provided by field H28 of the Hipparcos catalogue \citep{ESA-1997} and the correlation between Tycho-2 and Hipparcos proper motions, as encapsulated in the quantity 
$R$ of Table~1 of \citet{Hog-2000:b}, have both been been considered (see Jancart et al. 2006 for details). 
          
Since the above quantity follows a $\chi^2$ probability distribution function with 2 degrees of freedom, 
it is then possible
to compute the probability Prob$(\chi^2 > \chi^2_{\rm obs})$, giving the first kind risk of rejecting the null hypothesis 
$\vec{\mu}_{\rm Tycho} = \vec{\mu}_{\rm HIP}$ while it is actually true. This probability is listed in Table~\ref{Tab:SB},
along with $\chi^2_{\rm obs}$, and  the star is flagged as binary if Prob$ < 0.1$.

Only HIP~29740, HIP~76605 and HIP~97874 satisfy the test at the 10\% threshold. Note, however, that all the other stars
have parallaxes smaller than 5~mas, which make the test less efficient.

\section{Summary of the binary criteria and discussion}
\label{Sect:discussion}

The situation may be summarized as follows (see also last column of Table~\ref{Tab:SB}):
\begin{itemize}
\item Definite  binaries with known orbits:  HIP~4347, HIP~29740 (passes
all three binarity tests), HIP~43042, HIP~55852, CS~22942-019,
CS~22948-027;
\item Suspected binaries from astrometric data (either IAD or proper motions; no or inconclusive radial-velocity data):
HIP~34795,  HIP~76605, HIP~97874 (both astrometric tests yield positive results), HIP~107478;
\item Data inconclusive (mainly because of too small a parallax):
HIP~11595, HIP~25161, HIP~69834; 
\item Non-binary stars (mainly from radial-velocity data, not contradicted by astrometry): HIP~58596, BD~+3$^\circ$2688.
\end{itemize}    

The latter two non-binary stars are in fact good candidate
thermally-pulsing AGB stars, as revealed by their location in the HR
diagram (Table~\ref{Tab:H}  and right panel of Fig.~\ref{Fig:mdBa}).
Hence, they must not be binaries.

\begin{figure}
\resizebox{\hsize}{!}{\includegraphics{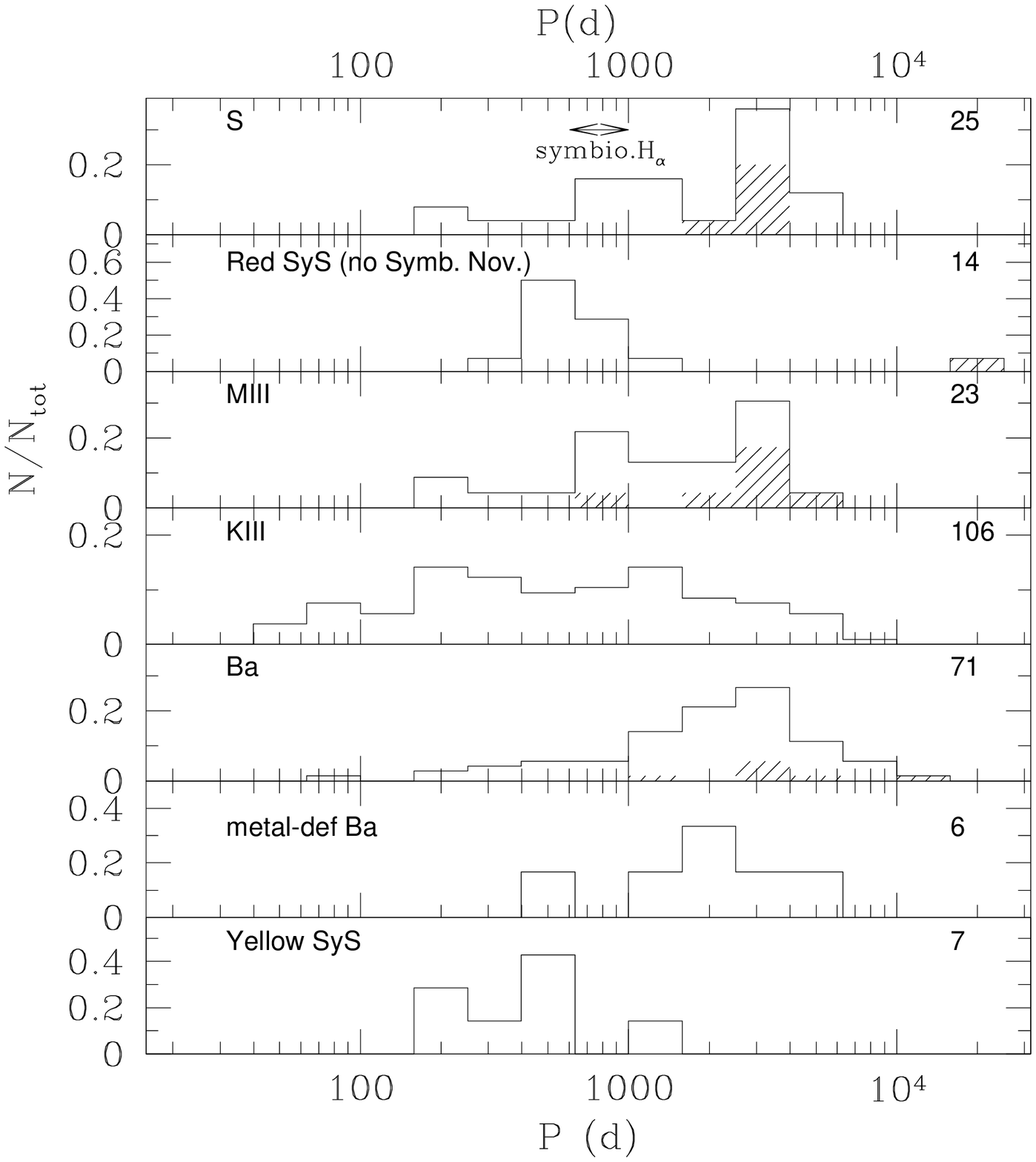}}
\caption[]{\label{Fig:P}
Comparison of the period distributions for samples of binary systems
with different kinds of red
giant primaries: S stars \protect\citep{Jorissen-VE-98}, red SyS  
\protect\citep[excluding symbiotic novae and symbiotic Miras; ][]{Muerset-99},
M giants  \protect\citep{Jorissen-2004:b} and 
K giants \protect\citep{Mermilliod-1996}. The lower two
panels  present the orbital-period distribution for barium stars
\protect\citep{Jorissen-VE-98} and
yellow SyS \protect\citep{Muerset-99}. In the S star panel, the arrow
marked H$_\alpha$ indicates the period range where binary S stars
exhibit H$_\alpha$ emission as a signature of weak symbiotic activity.
The shaded area marks stars with only a lower limit available on their orbital
period. The numbers in the upper right corner 
of each panel correspond to the sample size  
}
\end{figure}

Finally, we come to the central question of this paper: Why do the
metal-deficient barium stars, despite being binaries and occupying the same
location of the HR diagram as YSyS, do not exhibit symbiotic activity?
Three possible answers have been suggested in this paper:\\
(i) Some among the stars listed in Table~\ref{Tab:H} and displayed in Fig.~\ref{Fig:mdBa} are in fact not barium stars
(especially HIP~80843 = HD~148897).\\  (ii) Among those which are
barium stars, some seem to lie on the TP-AGB, and thus need not
be  binaries (HIP~58596 = HD~104340, BD~+3$^\circ$2688).  They
therefore cannot exhibit symbiotic activity.\\  (iii) Finally, there
remain a few genuine metal-deficient barium stars in the sample. Why are
they not symbiotic  stars?  It seems that the answer to that question
lies in the different period distributions for YSyS and metal-deficient
barium stars: YSyS  have shorter orbital period than  metal-deficient
barium stars, as seen in  Fig.~\ref{Fig:P}. This argument seems to apply
especially to HIP~29740 (=HD~43389), which has been assigned a very
bright absolute visual magnitude of $-3.5$ by the maximum likelihood
method of
\citet{Mennessier-97}. It is therefore expected to have a rather strong
mass loss rate, and be a good candidate YSyS. However, with its orbital
period of 1689~d, it lies at the long-period edge of the period
distribution of YSyS (Fig.~\ref{Fig:P}).   The same
difference seems to exist between the period distributions of red
symbiotics and binary S stars
\citep[][ and Fig.~\protect\ref{Fig:P}]{VanEck-Jorissen-02}. 

However,  a firm conclusion on
this issue should await the determination of the orbital periods for the
metal-deficient barium stars flagged as binaries by the IAD method (especially HIP~11595, HIP~25161, HIP~69834,
HIP~97874, HIP~107478), so as to make the comparison between the orbital period distributions of metal-deficient barium
stars and YSyS more meaningful.

\begin{acknowledgements}
This work was performed in the framework of the NATO Collaborative
Linkage Grant SA (PST.CLG.979128)6774/FP. We thank D. Pourbaix for the 
processing of the Hipparcos data of the metal-deficient stars. F. Carrier and
X. Bonfils are thanked 
for obtaining the Arcturus and HD~139409 spectra. 
LZ thanks I. Platais for valuable discussions and support.
The {\it Fonds National Suisse de la Recherche Scientifique} has funded the operations of the CORAVEL
spectrometer and of the Swiss 1-m telescope installed at the Haute-Provence
Observatory. 
\end{acknowledgements}

%
%
%



\def\aj{AJ}                   
\def\araa{ARA\&A}             
\def\apj{ApJ}                 
\def\apjl{ApJ}                
\def\apjs{ApJS}               
\def\ao{Appl.~Opt.}           
\def\apss{Ap\&SS}             
\def\aap{A\&A}                
\def\aapr{A\&A~Rev.}          
\def\aaps{A\&AS}              
\def\azh{AZh}                 
\def\baas{BAAS}               
\def\jrasc{JRASC}             
\def\memras{MmRAS}            
\def\mnras{MNRAS}             
\def\pra{Phys.~Rev.~A}        
\def\prb{Phys.~Rev.~B}        
\def\prc{Phys.~Rev.~C}        
\def\prd{Phys.~Rev.~D}        
\def\pre{Phys.~Rev.~E}        
\def\prl{Phys.~Rev.~Lett.}    
\def\pasp{PASP}               
\def\pasj{PASJ}               
\def\qjras{QJRAS}             
\def\sci{Science}             
\def\skytel{S\&T}             
\def\solphys{Sol.~Phys.}      
\def\sovast{Soviet~Ast.}      
\def\ssr{Space~Sci.~Rev.}     
\def\zap{ZAp}                 
\def\nat{Nature}              
\def\iaucirc{IAU~Circ.}       
\def\aplett{Astrophys.~Lett.} 
\def\apspr{Astrophys.~Space~Phys.~Res.}
\def\bain{Bull.~Astron.~Inst.~Netherlands} 
\def\fcp{Fund.~Cosmic~Phys.}  
\def\gca{Geochim.~Cosmochim.~Acta}   
\def\grl{Geophys.~Res.~Lett.} 
\def\jcp{J.~Chem.~Phys.}      
\def\jgr{J.~Geophys.~Res.}    
\def\jqsrt{J.~Quant.~Spec.~Radiat.~Transf.}
\def\memsai{Mem.~Soc.~Astron.~Italiana}
\def\nphysa{Nucl.~Phys.~A}   
\def\physrep{Phys.~Rep.}   
\def\physscr{Phys.~Scr}   
\def\planss{Planet.~Space~Sci.}   
\def\procspie{Proc.~SPIE}   

\let\astap=\aap
\let\apjlett=\apjl
\let\apjsupp=\apjs
\let\applopt=\ao

\bibliographystyle{apj}
\bibliography{/home/ajorisse/TEX/ajorisse_articles}

\end{document}